\documentclass[twoside]{article}

\usepackage[utf8]{inputenc}
\usepackage{fancyhdr}
\usepackage{natbib}
\usepackage[textwidth=6in,left=1.25in,right=1.25in,asymmetric]{geometry}
\usepackage{amsmath}
\usepackage{mathrsfs}

\usepackage{mathabx}

\usepackage{amsthm}

\usepackage{amsfonts}
\usepackage{amssymb}
\usepackage{graphicx}
\usepackage{placeins}
\usepackage{algorithm2e}
\usepackage{caption}
\usepackage{array}
\usepackage{multicol}
\usepackage{color}
\usepackage{mhequ}
\usepackage{url}
\usepackage{bbm}
\usepackage[font={small,it}]{caption}
\usepackage{stackrel}
\usepackage{enumerate}
\usepackage{datetime}
\usepackage{breqn}
\usepackage{authblk}

\theoremstyle{plain}

\newtheorem{theorem}{Theorem}[section]

\newtheorem{corollary}[theorem]{Corollary}

\theoremstyle{definition}
\newtheorem{definition}[theorem]{Definition}
\newtheorem{remark}{Remark}[section]

\renewcommand{\cite}{\citet}

\newcommand{\iid}{\mathop{\sim}\limits^{iid}  }

\newcommand{\gammad}{\text{Gamma}  }
\newcommand{\betad}{\text{Beta}  }

\newcommand{\discreted}{\text{Discrete}}
\newcommand{\bx}{\mathbf{x}}

\newcommand{\bb}[1]{\mathbb{#1}}
\newcommand{\EX}[1]{E \left( #1 \right)}
\newcommand{\EXc}[1]{E \left\{ #1 \right\}}

\newcommand{\Var}[1]{\mathrm{var} \left( #1 \right)}
\newcommand{\Varc}[1]{\mathrm{var} \left\{ #1 \right\}}
\newcommand{\Vars}[1]{\mathrm{var} \left[ #1 \right]}

\newcommand{\PR}[1]{\mathrm{pr} \left( #1 \right)}

\newcommand{\mc}[1]{\mathcal{#1}}
\newcommand{\bone}[1]{\mathbbm{1}\left\{ #1 \right\}}

\newcommand{\bigO}[1]{\mc O \left( #1 \right)}
\newcommand{\bigOc}[1]{\mc O \left\{ #1 \right\}}
\newcommand{\bigOs}[1]{\mc O \left[ #1 \right]}

\DeclareMathOperator{\argmin}{argmin}

\newcommand{\Bet}[2]{\mathrm{Beta}\left( #1, #2 \right)}
\newcommand{\Gam}[2]{\mathrm{Gamma}\left( #1, #2 \right)}
\newcommand{\Binom}[2]{\mathrm{Binom} \left( #1, #2 \right)}
\newcommand{\be}{\begin{equs}}
\newcommand{\ee}{\end{equs}}

\makeatletter
\providecommand*{\input@path}{}
\g@addto@macro\input@path{{./}{../input/}}
\makeatother

\title{Estimating the observable population size from biased samples: a new approach to population estimation with capture heterogeneity}

\author[1]{James E. Johndrow \thanks{jj@stat.duke.edu}}
\author[2]{Kristian Lum \thanks{kl@hrdag.org}}
\author[3]{Daniel Manrique-Vallier \thanks{dmanriqu@indiana.edu} \thanks{Authors contributed equally and are listed alphabetically}}
\affil[1]{Department of Statistical Science, Duke University}
\affil[2]{Human Rights Data Analysis Group}
\affil[3]{Department of Statistics, Indiana University}
\date{\today}
\allowdisplaybreaks
\begin{document}
\maketitle

\begin{abstract}
Capture-recapture methods aim to estimate the size of a closed population on the basis of multiple incomplete enumerations of individuals. In many applications, the individual probability of being recorded is heterogeneous in the population. Previous studies have suggested that it is not possible to reliably estimate the total population size when capture heterogeneity exists. Here we approach population estimation in the presence of capture heterogeneity as a latent length biased nonparametric density estimation problem on the unit interval. We show that in this setting it is generally impossible to estimate the density on the entire unit interval in finite samples, and that estimators of the population size have high and sometimes unbounded risk when the density has significant mass near zero.  As an alternative, we propose estimating the population of individuals with capture probability exceeding some threshold. We provide methods for selecting an appropriate threshold, and show that this approach results in estimators with substantially lower risk than estimators of the total population size, with correspondingly smaller uncertainty, even when the parameter of interest is the total population. The alternative paradigm is demonstrated in extensive simulation studies and an application to snowshoe hare multiple recapture data. 
\end{abstract}

\section{Introduction} \label{sec:intro}
Capture-recapture is a class of statistical methods designed to estimate the size of a closed population. The data required for this task are $T$ incomplete but non-disjoint samples of individuals from the population. Each individual is identified across all lists to produce a ``capture history", a length $T$ binary vector giving the pattern of presence and absence on the lists. A joint model for multivariate binary outcomes is then fit to the capture histories of all individuals who appeared in the sample, and used to estimate the number of individuals not captured on any list. Crucially, in order to produce a useful estimate of the total population size $N$, the model must explain not only the observable data, but also the distribution of individuals not observed on any list. Since the model can only be fit to observed individuals, the quality of inference will depend heavily on the relationship between the observed and the non-observed individuals implied by the joint model.

The literature in capture-recapture methods has long recognized the need for specialized methods to deal with populations in which capture probabilities vary from individual to individual. This phenomenon is referred to as capture heterogeneity. In the absence of covariate information that can control for the differences in observability through, for instance, stratified estimation; see \citet{sekar:deming:1944}, most proposed approaches rely on joint sampling models that include some form of individual random effects. The random effects model variation across individuals in the probability of being observed. Early proposals modeled the random effects using a parametric mixing distribution, often chosen because of its mathematical tractability (\citet{Sanathanan1973, Agresti1994, Darroch1993}). Other approaches have sought to estimate the mixing distribution in a non-parametric way (\citet{mao:2008:cr:npmle:computing}). 

The presence of capture heterogeneity is equivalent to bias in the sampling process, since individuals with smaller values of the random effect are necessarily underrepresented on the partial lists.  Although not characterized as resulting from sampling bias, this situation was nonetheless recognized in a series of articles, notably by \citet{huggins:2001:cr:heterogeneity:identif} and \citet{Link2003}, that discussed the identifiability of $N$ in heterogeneous capture-recapture. In particular \citet{Link2003} showed how simple parametric models produced almost identical fitted observed frequencies, while inducing completely different estimates for the number of unobserved individuals. Thus, while using parametric forms for the random effect distribution addresses the immediate issue of identifiability, when the model is misspecified, the resulting population estimates can be wildly inaccurate. This has led some authors to suggest estimating a lower bound on the total population size that is valid for parametric models satisfying certain conditions (for example, see \citet{chao1987estimating}).

In this article, we show that population estimation with capture heterogeneity is analogous to a latent density estimation problem in the presence of length bias. Length, or size, bias is a type of sampling bias where the probability of observing data is proportional to the magnitude of the observation, as outlined in \cite{patil1977weighted, patil1978weighted}. There is an extensive literature on density estimation in the presence of length bias, including \cite{vardi1985empirical}, \cite{vardi1982nonparametric}, \cite{asgharian2002length}, \cite{de2004nonparametric}, \cite{jones1991kernel}, and \cite{gill1988large}, among others. Curiously, the capture-recapture literature and the sampling bias literature are almost completely orthogonal, possibly because the canonical approach is to consider the individual list capture probabilities rather than the probability of capture \emph{on at least one list}, as we do here. The only reference to length bias in the capture-recapture literature that we are aware of is \cite{chen2002estimation}, which is a very different setting from ours, requires the availability of individual covariates, and makes the assumption that individual capture probabilities are bounded away from zero. Therefore, a major contribution of this paper is to bring these two literatures closer by making explicit this connection and providing theoretical results on the risk of capture-recapture estimates in the presence of capture heterogeneity that are a function of nonparametric density estimates under length bias sampling. A secondary contribution is to the literature on Bayesian nonparametric methods for biased sampling, which is relatively thin, see \cite{hatjispyros2015bayesian} and \cite{kunihama2014nonparametric}. In this context, our method can be viewed as a Bayesian missing data approach to nonparametric density estimation under length bias.

Our second major contribution is to argue that, while this intrinsic limitation of capture-recapture methods in the presence of capture heterogeneity makes the problem of estimating $N$ essentially ill-posed, there are other meaningful, interpretable quanities related to total population size that can reliably be estimated in this scenario. We thus propose approaching capture-recapture estimation from a different perspective. Instead of estimating the total population size, we advocate estimating the \emph{observable} population size, which, informally, is the size of the total population that has non-negligible probability of being observed. To make this approach rigorous, we define the concept of $\alpha$-observable populations, the population of individuals with probability at least $\alpha$ of being observed on at least one list. We show that, in addition to being properly defined quantities with an intuitively sound interpretation, estimators of the $\alpha$-observable population size have superior properties to estimators of $N$. Moreover, they often have lower risk \emph{as estimators of $N$} than the corresponding unbiased estimator of $N$. We propose a Bayesian approach to inference based on discrete mixtures, and show how this approach is ideally suited to estimating the size of $\alpha$-observable populations. We also outline a strategy for choosing the optimal value of $\alpha$ as a byproduct of model fitting. 

\section{Heterogeneity in Capture Recapture}
Consider a sample of $m$ individuals captured or recorded from a population of unknown size $N$ during $T$ capture attempts. We represent each individual's {\it capture history} using binary indicators, $x_{it}$ for $i=1, ..., m$ and $t=1, ..., T$, which take the value 1 if $i$th individual appeared on the $t$th list and 0 otherwise. These data can be summarized by a $2^T$ contingency table, with cell counts $n(x)$ for $x \in \{0,1\}^T$ representing the number of individuals in the sample with observed capture history $x$. For example, $n\{(010)\}$ is the number of individuals captured on the second list, but not on the first or third. The count in the cell indexed by all zeroes, $n(\zeta)$ where $\zeta$ is the zero vector, is unobservable by definition, as it represents the total number of individuals in the population that were not recorded on any list. This value is the object of inference, since adding it to $m$ gives an estimate of the total population size $N$. 

Capture heterogeneity refers to the case when the probability of capture varies across individuals in the population, resulting in a sample that is biased toward individuals with high capture probability. In our initial development, we abstract from the multiple list structure of the data, and consider only the individual probability of being observed on at least one list. We will demonstrate that knowledge of the distribution of this parameter is sufficient to estimate $N$, and consider a natural estimator. 

Let $\pi_{x,i}$ be the probability that individual $i$ has capture history $x$, and $p_i = 1-\pi_{\zeta,i}$ the probability of appearing in least one list. In the absence of heterogeneity $p_i = p$ for all $i$. For known $p$, the estimator $\widehat{N} = m p^{-1}$ is a commonly used estimator of $N$. When $P \sim F$ for a probability distribution $F$ supported on the unit interval, an analogous estimator is 
\be
m \left\{\int_{[0,1]} p \, F(dp) \right\}^{-1} = m \{E_F(P)\}^{-1};
\ee 
when $F$ has a density $f$, we will sometimes write $E_f(P)$ instead of $E_F(P)$. The following remark shows that it is asymptotically equal to a regularized maximum likelihood estimate, where the regularization can be interpreted as a continuity correction for the parameter $N$. 
\begin{remark} \label{rem:nhatnonhet}
Suppose $p < 1$. Consider the penalized Binomial log-likelihood with $N$ treated as a continuous parameter
\be
\tilde{\ell} ( m \mid N,p) &= \log \left\{ \frac{\Gamma(N+1)}{\Gamma(m+1) \Gamma(N-m+1)} \right\}  + m \log(p) + (N-m) \log (p) - \frac{1}{2} \log \left( \frac{N}{N-m} \right). \label{eq:penlik}
\ee
Asymptotically, $\tilde{\ell} (m \mid N,p)$ is equal to a $\text{Binomial}(N-1/2,p)$ log-likelihood, so the penalty is analogous to a continuity correction. Moreover, $\widehat{N} = \argmin_N \tilde{\ell} (m \mid N,p)$ satisfies
\be
\widehat{N} = m p^{-1} + \bigO{m^{-1}}. \label{eq:nhat}
\ee
More generally if $P \sim F$ for a discrete measure with finitely many atoms, then $m \sim \textnormal{Binomial}\left\{ N, E_F(P) \right\}$, and $\widehat{N} = \argmin_N \tilde{\ell} (m \mid N,p)$ satisfies
\be
\widehat{N} \mid m,f  &= m \{E_F(P)\}^{-1} + \bigO{m^{-1}}.
\ee
This suggests the estimator $m \{ E_F(P) \}^{-1}$ for general $F$. So in the presence of capture heterogeneity, nonparametric estimation of $N$ and $n(\zeta)$ can be equated with estimation of the \emph{expectation} of $P$ with respect to $F$, and the associated estimator of $N$ is $m$ divided by $E_F(P)$.  
\end{remark}

\begin{figure}[h]
\centering
\includegraphics[width=0.9\textwidth]{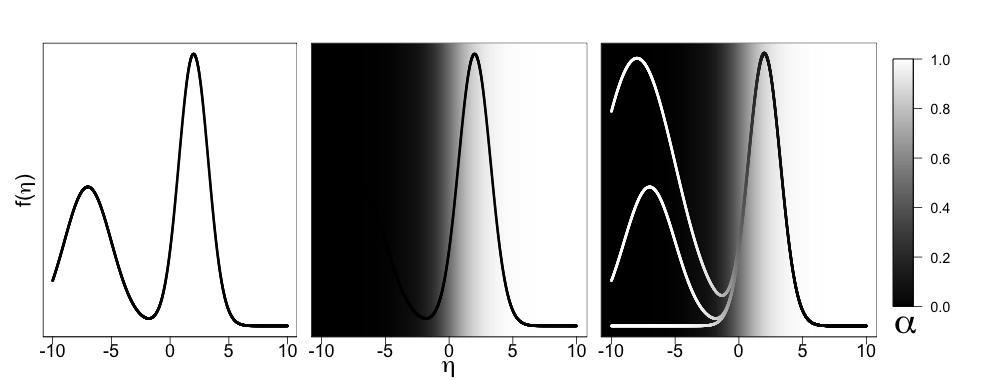}
\caption{Left panel: a generic density $f^*(\eta)$. Center panel: same density as in left panel, but with background shading proportional to $\varphi^{-1}(\eta)$ with $\varphi$ the logit function. Right panel: several possible densities that share the same shape in the ``observable'' region, but different shapes in the ``unobservable'' region.    } \label{fig:cartoon}
\end{figure}

The problem of nonparametric estimation of $F$ differs from typical distribution estimation problems. For simplicity, suppose $F$ has density $f$. In the traditional density estimation setting, data in the neighborhood around the value $p$ appears in the sample in proportion to $f(p)$. 
In contrast, in the capture-recapture setting, the probability of observing data in a small neighborhood around $p$ is proportional to $p f(p)$. This type of sampling bias is referred to as length bias (\cite{patil1977weighted}). In this setting, the probability of observing data in the region near zero in finite samples is vanishingly small for bounded densities $f(p)$, even if $f$ has significant mass near zero. Figure \ref{fig:cartoon} shows a graphical representation of this phenomenon, with the density represented on the logit scale for clarity. Specifically,  $\eta = \varphi(p)$, where $\varphi : [0,1] \to \bb R$ is the logit function.   The left panel represents the ``true" density $f^*(\eta)$. The center uses shading to represent \emph{observability} $\varphi^{-1}(\eta)$ of the individuals with random effects in that region; individuals in the dark region are unlikely to be observed. As a result we have the situation depicted in the right panel: regardless of the true density, the three densities shown here fit the observable data equally well, but lead to a dramatically different characterization of the unobserved population.

\section{Estimation of $\alpha$-observable populations} \label{sec:aobsprops}
We now introduce an alternative inferential approach that estimates the $\alpha$-observable population size $N_{\alpha}$ rather than the total population size $N$. Definition \ref{def:alphaob} formalizes the concept of $\alpha$-observability.

\begin{definition} \label{def:alphaob}
Individual $i$ is $\alpha$-observable if $1-\pi_{\zeta, i} = p_i > \alpha$. We define the \textit{$\alpha$-observable population size}, $N_{\alpha}$, as $N_{\alpha} = \sum_{i=1}^N \bone{p_i > \alpha}$. We call $[\alpha,1]$ the \textit{$\alpha$-observable region}. Any set such that $A \subset [\alpha,1]$ is an \textit{$\alpha$-observable set}. Any set that is not $\alpha$-observable is \textit{$\alpha$-unobservable}.
\end{definition}

In contrast to estimating $N$, estimating $N_{\alpha}$ requires learning the distribution $F$ only on $[\alpha,1]$. Then the analogue of $\widehat N$ is 
\be
\widehat{N}_{\alpha} = m_{\alpha} \{1-F(\alpha)\} \{E_F(P)\}^{-1} = m_{\alpha} \{E_{F_{\alpha}}(P)\}^{-1}, 
\ee
where $m_{\alpha}$ is the number of observed individuals having $p \in \bb [\alpha,1]$ and $E_{F_{\alpha}}(P) = E_F(P \mid P > \alpha)$. So to estimate $\widehat{N}_{\alpha}$ we only need to integrate over $[\alpha,1]$, not the entire unit interval.

\subsection{Risk of estimators of $N$ and $N_{\alpha}$} \label{sec:estnalpha}
We now show that estimators of $N_{\alpha}$ have lower risk than estimators of $N$. Intuitively, low risk of any nonparametric estimator $\widehat{F}$ of $F$ requires that we observe data with high probability in sets whose $F$ measure is bounded away from zero. We refer to these sets as \emph{non-negligible}. 

\begin{definition}[Non-negligible sets]
For any $\epsilon > 0$, a Borel set $A$ is non-negligible with respect to $F$ and $\epsilon$ if $\int_{p \in A} F(dp) > \epsilon$. 
\end{definition} 

Theorem \ref{thm:hpregion} shows that we can guarantee a uniform minimal probability of observing data in non-negligible $\alpha$-observable sets in finite samples, but not in all non-negligible sets. Therefore, nonparametric estimators of $N_{\alpha}$ are expected to have better finite-sample performance than estimators of $N$. All proofs are deferred to the Appendix.

\begin{theorem}[Finite sample bounds in $\alpha$-observable populations] \label{thm:hpregion}
Suppose the true population size is $N$ and fix $\epsilon > 0$. Let $\mc A$ be the set of all $(F,\epsilon)$ non-negligible $\alpha$-observable intervals, and for any $A \subset \bb [0,1]$, let $M(A) := \sum_{i : p_i \in A} \bone{\sum_{t=1}^T x_{it}>0}$, the number of observations with $p \in A$. Then 
\begin{align}
\inf_{A \in \mc A} \mathrm{pr} \{M(A) > 0\} = 1-(1-\alpha \epsilon)^N . \label{eq:lbprob}
\end{align}
\end{theorem}

That is, for fixed $N$, the probability that {\it some data} are observed in any non-negligible set is bounded below by the right hand side of \eqref{eq:lbprob}, which is a function of $\alpha$. For $\alpha=0$, the only uniform bound is the trivial bound $\inf_{A \in \mc A} \mathrm{pr} \{M(A) > 0\} = 0$. Another implication of Theorem \ref{thm:hpregion} is that for fixed $\alpha$ and $\epsilon$ and any $q \in (0,1)$, there exists $N<\infty$ large enough that the probability of observing data in $A \in \mc A$ is uniformly bounded below by $q$. Of course, no finite value of $N$ would be sufficient to guarantee a uniform lower bound on the probability of observing data in an arbitrary non-negligible subset of the unit interval. Finally, Theorem \ref{thm:hpregion} suggests one approach for choosing values of $\alpha$ based on the number of unique observed individuals $m$. Although $N$ is unknown, it is always the case that $N \ge m$, so the probability of observing data in non-negligible $\alpha$-observable sets can be lower-bounded by (\ref{eq:lbprob}) with $N=m$. Therefore, one way to assess different choices of $\alpha$ is to compute (\ref{eq:lbprob}) for the observed population size $m$ and a range of $\epsilon$ values. 

Now we consider the risk of estimators of $N$ and $N_{\alpha}$. Throughout, we will assume that we observe directly length-biased data proportional to $p F(dp)$. In capture-recapture applications, $p$ is latent, so the results in this section provide a lower bound on the risk of capture-recapture estimates of $N$. The distribution of the observed data is 
\be
G(dp) = \frac{p F(dp)}{E_f(P)}, \label{eq:obsdens}
\ee
since $G(dp)$ must integrate to 1. Consider an estimator $\widehat{G}$ of $G$. We propose to use
\be
\widehat{F}(dp) = \frac{p^{-1} \widehat{G}(dp)}{E_{\widehat{G}}(P^{-1}) } \label{eq:fhat}
\ee
as an estimator of $F$. This is similar to the approach in \cite{hatjispyros2015bayesian}. This estimator is appropriate in at least one sense, which is to say that since $F(dp) \propto p^{-1}\, G(dp)$, it must be the case that
\be
F(dp) = \frac{p^{-1} G(dp)}{ E_G(P^{-1})} = p^{-1} G(dp) E_f(P),
\ee
so $E_G(P^{-1}) =  \{E_F(P)\}^{-1}$, and $\widehat{F}$ is the plug-in estimator of $F$ derived from any estimator of the observable distribution $\widehat{G}$.  

We now consider two procedures for estimation of $\widehat{G}$. Critically, when $F$ is unknown, we cannot estimate $E_F(P)$ without implicitly choosing an estimator $\widehat{F}$ because of the length bias in the observed data. We consider two possible estimators of $\widehat{G}$: the empirical measure and a histogram estimator. These estimators are classical analogues of the Bayesian Dirichlet process model we propose in \S \ref{sec:dp}. Since $E_G(P^{-1}) =  \{E_F(P)\}^{-1}$, in either case the estimator $\widehat{N}$ is given by $m E_{\widehat{G}}(P^{-1})$. When $\widehat{G}$ is the empirical measure, this is just the sample mean of $p^{-1}$, $\widehat{N} = \sum_{i=1}^m p_i^{-1}$, and we have the following result for the mean squared error of $\widehat{N}$. 

\begin{theorem}[Mean squared error for empirical measure estimator] \label{thm:empmeasure}
If $\widehat{G}(dp) = \sum_{i=1}^m \delta_{p_i}(dp)$ then the mean squared error $\Delta^2(\widehat{N},N)$ is given by
\be
\Delta^2(\widehat{N},N) = N \{ E_F(P^{-1}) - 1 \}. \label{eq:empmeasmse}
\ee 
\end{theorem}
It is immediate that for $\Delta^2(\widehat{N},N)<\infty$, we need $E_F(P^{-1}) < \infty$. In many seemingly mundane cases this fails, for example when $F$ is a $\text{Beta}(a,b)$ distribution with $a \le 1$, which includes the uniform distribution on the unit interval. The following Corollary shows that the empirical measure estimator of $N_{\alpha}$ always has finite $L_2$ risk, and the risk decreases with increasing $\alpha$.
\begin{corollary}[Mean squared error for empirical measure estimator of $N_{\alpha}$] \label{cor:nalphaemp}
For any $\alpha \in (0,1]$ if $\widehat{N}_{\alpha}$ is the empirical measure estimator of $N_{\alpha}$, then
\be
\Delta^2(\widehat{N}_{\alpha},N_{\alpha}) = N_{\alpha} \{ E_{F_{\alpha}}(P^{-1}) - 1 \} \le N_{\alpha} (\alpha^{-1}-1),
\ee
and $N_{\alpha}^{-1} \Delta^2(\widehat{N}_{\alpha},N_{\alpha})$ is monotone nonincreasing in $\alpha$.
\end{corollary}
Finally, in many cases $\widehat{N}_{\alpha}$ using the empirical measure estimator of $F$ will have lower $L_2$ risk for $N$ than the corresponding empirical measure estimator of $N$, and $\widehat{N}$ cannot be a minimax estimator of $N$ in the nonparametric regime.
\begin{corollary}[Mean squared error for estimation of $N$ by the empirical measure estimator of $N_{\alpha}$] \label{cor:msenalphan}
For any $\alpha \in (0,1]$ if $\widehat{N}_{\alpha}$ is the empirical measure estimator of $N_{\alpha}$, then
\be
\Delta^2(\widehat{N}_{\alpha},N) = N_{\alpha} \{ E_{F_{\alpha}}(P^{-1}) - 1 \} + (N-N_{\alpha})^2 \le N_{\alpha} (\alpha^{-1}-1) + (N-N_{\alpha})^2.
\ee
Therefore, $\Delta^2(\widehat{N}_{\alpha},N) < \Delta^2(\widehat{N},N)$ whenever $N_{\alpha} \{ E_{F_{\alpha}}(P^{-1}) - 1 \} + (N-N_{\alpha})^2 < N \{ E_F(P^{-1}) - 1 \}$. In particular, $\widehat{N}$ cannot be a minimax estimator of $N$ when $F$ is unknown, since there exists $F$ for which $\Delta^2(\widehat{N},N) = \infty$, while $\Delta^2(\widehat{N}_{\alpha},N) < \infty$ for all $F$.
\end{corollary}

We now consider the case where $F$ has a density $f$ and $\widehat{g}$ is the histogram estimator. The histogram estimator is a commonly used nonparametric density estimator. We have the following result for the risk of histogram-based estimators of $N$. 
\begin{theorem}[Mean squared error of histogram estimators of $N$] \label{thm:histmse}
Suppose $\widehat{g}$ is the histogram estimator with bin width $h$. Assume $f$ is twice continuously differentiable. Then the mean squared error $\Delta^2(\widehat{N},N)$ satisfies
\be 
0 \le \Delta^2(\widehat{N},N) - N   \left[ E_f(P^{-1}) - \{E_f(P)\}^{-1} \right] + \bigO{Nh}  \le \frac{N^2 h^2}{4} \left\{ E_f(P^{-1}) + E_f\{|f'(P)|\} \right\}^2 \label{eq:mse}
\ee 
Further, the asymptotic variance (as $n \to \infty$ and $h \to 0$) of $\widehat{N}$ is given by $N   \left[ E_f(P^{-1}) - \{E_f(P)\}^{-1} \right]$, and the asymptotic bias is bounded above by $(N^2 h^2/4) \left\{ E_f(P^{-1}) + E_f\{|f'(P)|\} \right\}^2$. 
\end{theorem}

From equation \eqref{eq:mse}, it is clear that when $h$ shrinks faster than $N^{-1/2}$, the asymptotic mean squared error and asymptotic variance are identical. Moreover, for twice differentiable densities $f$, we can improve upon the asymptotic mean squared error for the empirical measure estimator, since $\{ E_F(P^{-1}) - 1 \} \ge [ E_f(P^{-1}) - \{E_f(P)\}^{-1} ]$. Nonetheless, $\Delta^2(\widehat{N},N) < \infty$ still requires $E_f(P^{-1}) < \infty$. The following Corollary shows that estimating $N_{\alpha}$ has similar benefits when we use a histogram estimator of $g$. 

\begin{corollary}[Mean squared error of histogram estimators of $N_{\alpha}$] \label{cor:nalphamse}
Under the conditions of Theorem \ref{thm:histmse}, if $\alpha \in (0,1]$ and $h = N^{-(1+\delta)/2}$ for $\delta > 0$ then
\be
\lim_{N_{\alpha} \to \infty} N_{\alpha}^{-1} \Delta^2(\widehat{N}_{\alpha},N_{\alpha}) = \left[ E_{f_{\alpha}}(P^{-1}) - \{E_{f_{\alpha}}(P)\}^{-1} \right] < \alpha^{-1}.
\ee
Furthermore, if $E_{f_{\alpha}}(P^{-1}) - \{E_{f_{\alpha}}(P)\}^{-1}$ is monotone nonincreasing in $\alpha$, then the asymptotic mean squared error of $\widehat{N}_{\alpha}$ is also monotone nonincreasing in $\alpha$. 
\end{corollary}
We conjecture that $E_{f_{\alpha}}(P^{-1}) - \{E_{f_{\alpha}}(P)\}^{-1}$ is monotone nonincreasing in general, though something stronger than the obvious convexity argument is required to show this result. Empirically, this holds for all of the distributions we have tested. We also have a result similar to Corollary \ref{cor:msenalphan} for the histogram estimator.
\begin{corollary}[Mean squared error for estimation of $N$ by histogram estimator of $\widehat{N}_{\alpha}$] \label{cor:msenalphanhist}
For any $\alpha \in (0,1]$ if $\widehat{N}_{\alpha}$ is the histogram estimator of $N_{\alpha}$ and the conditions of Theorem \ref{thm:histmse} hold, then
\be
\Delta^2(\widehat{N}_{\alpha},N) &\le \Delta^2(\widehat{N}_{\alpha},N_{\alpha}) + (N-N_{\alpha})^2 < \infty
\ee
It follows that $\widehat{N}$ cannot be a minimax estimator of $N$ when $F$ is unknown, since there exists $F$ for which $\Delta^2(\widehat{N},N) = \infty$, while $\Delta^2(\widehat{N}_{\alpha},N) < \infty$ for all $F$.
\end{corollary}

Theorems \ref{thm:empmeasure} and \ref{thm:histmse} show that in many cases, estimators of $N$ have unbounded risk, while Corollaries \ref{cor:nalphaemp} and \ref{cor:nalphamse} guarantee that for any $\alpha > 0$, there exist estimators of $N_{\alpha}$ with finite risk for any $F$. Further, these results imply that the choice of $\alpha$ can be viewed as a tradeoff between higher mean squared error and estimating a quantity closer to the total population size. Finally, Corollaries \ref{cor:msenalphan} and \ref{cor:msenalphanhist} show that estimators of $N_{\alpha}$ often have lower risk \emph{for estimation of} $N$ than the corresponding estimators of $N$ and are superior in the minimax sense. These are the basic lessons that motivates the methods we propose for estimation of $N_{\alpha}$ and choice of $\alpha$ in the sequel. 

\section{Model-based estimation of $N_{\alpha}$}
Thus far, the discussion has centered on $p_i = 1-\pi_{\zeta,i}$ without specifying a model for the other cell probabilities. In capture-recapture, we must use the data on presence or absence on the $T$ lists to estimate the latent distribution $F$ and the population size $N$. This is generally done by specifying a model on the individual-level cell probabilities $\pi_{x,i}$, which induces a model on $p_i = 1-\pi_{\zeta,i}$, and on $F$ by marginalizing over individuals. We now restrict our attention to a specific class of such models, the $M^{th}$ class. In modeling, it is common to parametrize probabilities via a monotone nondecreasing transformation $\varphi^{-1} : \bb R \to [0,1]$, such as the logit or probit function. We will follow this convention for the remainder of the paper, and will write $\eta = \varphi(p)$ for the transformed individual observabilities. Densities and distribution functions supported on the real line induced by the transformation of $P \sim F$ via $\varphi$ are represented with superscript $^*$, for example, $f^*(\eta)$ and $F^*$. 

\subsection{$M^{th}$ Model}
The $M^{th}$ model (\cite{Agresti1994}) is given by the following joint distribution for the list capture random variables $X_1,\ldots,X_T$:
\be
\PR{X_{i1}=x_1,\ldots,X_{iT}=x_T \mid \theta, \beta} &= \prod_{t=1}^T \varphi^{-1}(\theta_i + \beta_t)^{x_{it}}  \{1-\varphi^{-1}(\theta_i + \beta_t)\}^{1-x_{it}}, \quad \theta_i \stackrel{iid}{\sim} G^*, \label{eq:mod}
\ee
where the $\theta_i$ are individual-specific observability effects distributed according to a distribution $G^*$ supported on $\bb R$, and the $\beta_t$ are global list effects. By \eqref{eq:mod}, individual $i$ is $\alpha$-observable in this model if 
\be
p_i= \varphi^{-1}(\eta_i) = 1-\prod_{t=1}^T \{1-\varphi^{-1}(\theta_i + \beta_t)\} > \alpha. \label{eq:aobsmth}
\ee
Clearly, the distribution $G^*$ on $\theta$ induces a distribution $F^*$ for $\eta = \varphi(p)$, the individual observabilities under the transformation $\varphi$. Thus, informally speaking, the parameter $\theta$ is analogous to the individual observability $\eta$. Some additional information on $\alpha$-observability and the induced distribution on $\eta$ are provided in Supplementary Materials. 

Historically, the literature has focused on certain parametric choices for $G^*$ that lead to computational tractability. An extensive review is provided in Supplementary materials. To summarize, parametric choices for $G^*$ perform very poorly under model misspecification, since any parametric choice for $G^*$ implies a specific shape for the distribution in the unobservable region. As such, we focus on nonparametric estimation of $G^*$, consistent with our discussion thus far and with the work of \cite{mao:2008:cr:npmle:computing}. In particular, we propose estimators of $N_{\alpha}$ and $N$ in the context of a Bayesian nonparametric $M^{th}$ model, and compare empirical performance to estimation of $N_{\alpha}$ and $N$ using parametric choices of $G^*$.

\subsection{A Bayesian nonparametric prior for $G^*$} \label{sec:dp}
Motivated by the desire to estimate $N_{\alpha}$ using a flexible nonparametric model for $G^*$ in the $M^{th}$ model, we propose the Dirichlet process prior for $G^*$. Mixture models for $G^*$ have been previously proposed by \cite{pledger2000unified}. \cite{ford2015modelling} suggests a Dirichlet process model in the context of capture-recapture, but does not apply the model to population estimation. Conversely, our focus is on population estimation, and particularly on the properties of the model with respect to estimation of $N_{\alpha}$ and an analogue of the lower-bound estimate of \cite{chao1987estimating}. We take a Bayesian approach to inference, and provide a Markov chain Monte Carlo algorithm for computation.

The Dirichlet process is a stochastic process specified by a concentration parameter $\alpha$ and a base measure $H$. It is a random probability measure $\mu(\cdot)$ that can be characterized by its stick-breaking representation 
\begin{align}
\mu(A) &= \sum_{h=1}^{\infty} \nu_h \delta_{\theta^*_h}(A), \quad \theta^*_h \stackrel{iid}{\sim} H, \label{eq:dp} \\
\nu_h &= \nu^*_h \prod_{l < h} (1-\nu^*_l), \quad \nu^*_1,\nu^*_2,\ldots \stackrel{iid}{\sim} \Bet{1}{\alpha_0}, \label{eq:dpstick}
\end{align} 
for $H$-measurable sets $A$, where $\delta_{\theta^*}$ is a Dirac measure at $\theta^*$. This model can also be represented hierarchically using a countably supported latent class variable $z$
\be
\mu(A \mid z) &= \delta_{\theta^*_z}(A), \quad \PR{z = h} = \nu_h, \label{eq:dphier}
\ee
a representation that is commonly employed for computation. When a random probability distribution $G^*$ can be characterized by (\ref{eq:dp})-(\ref{eq:dpstick}), we write $G^* \sim DP(\alpha_0,H)$. From (\ref{eq:dp}), it is clear that the Dirichlet process with normal base measure is analogous to the empirical measure estimator, in that both are atomic. If the base measure is instead supported on a function space, then the resulting estimator would be analogous to a histogram or kernel density estimator. Thus, the discussion in Section \ref{sec:estnalpha} is for the most part relevant to the behavior of the Dirichlet process as a prior for $G^*$ in the $M^{th}$ model.

To complete a Bayesian specification for the likelihood in \eqref{eq:mod} with Dirichlet process prior on $G^*$, we choose a Gaussian base measure and Gamma prior on $\alpha_0$, giving
\be
G^* &\sim DP\{\alpha_0,N(0,\sigma^2_{G^*})\}, \quad \beta_t \sim N(0,\sigma^2_{\beta}), \quad \alpha_0 \sim \Gam{a}{b}. \label{eq:dpmod}
\ee

The induced prior on $F^*$, the distribution of $\eta$, is also a random measure. In particular, if $\varphi$ is the logit function, $\beta_t = \beta$ for all $t$, and $\mu_{G^*} = \sum_{h=1}^{\infty} \nu_h \delta_{\theta_h}$ is the probability measure corresponding to $G^*$, then $\mu_{F^*} = \sum_{h=1}^{\infty} \nu_h \delta_{\eta(\theta,\beta)}$ where $\mu_{F^*}$ is the probability measure corresponding to $F^*$. We use $\varphi=\Phi$, the probit function, which facilitates computation. We describe an efficient Markov chain Monte Carlo algorithm based on the truncation approximation of \cite{ishwaran2001gibbs} in Supplementary Materials.

Finally, we specify the prior $p(N) \propto N^{-1} \bone{N \in \bb N, N \ge m}$ for the unknown population size. The resulting conditional distribution of $N-m$ given $\theta^*_{[1:K]}$, $\beta_{[1:T]}$, and $\nu_{[1:K]}$ for a $K$-length truncation of the stick-breaking process is 
\be
p(N-m \mid \theta^*_{[1:K]}, \beta_{[1:T]}, \nu_{[1:K]} ) &\sim \text{NegBin}\left(m, 1-\sum_{h=1}^K \nu_h \prod_{t=1}^T \Phi(-\theta^*_h - \beta_t) \right) \\
&\sim \text{NegBin}\left\{ m, 1-E_{F}(P) \right\}.
\ee
It follows that 
\be
E(N \mid \theta^*_{[1:K]}, \beta_{[1:T]}, \nu_{[1:K]} ) &= m+ \frac{m \{1-E_{F}(P)\}}{E_{F}(P)} = \frac{m}{E_{F}(P)},
\ee
so the posterior mean of $N$ in this model is equivalent to the point estimates of the population size considered in \S \ref{sec:aobsprops}, where the estimator for $F$ is induced by the Dirichlet process prior on $G^*$ via the transformation in \eqref{eq:aobsmth}.

\subsection{Estimation of $N_{\alpha}$ and choice of $\alpha$}
At each iteration the Markov chain Monte Carlo algorithm obtains samples of latent class variables $z_i$ for every individual, as defined by a truncation of \eqref{eq:dphier}. Via these samples, every individual is assigned a value of $\theta_i$. In addition, the fully Bayesian approach results in sampling of \emph{unobserved} members of the population conditional on the sampled value of $N$. Thus, at every iteration of the sampler, every member of the population, including unobserved members, is assigned a value of $\theta_i$. 

Conditional on samples $N^{(s)}$ of $N$, $\theta^{(s)}_i$ of $\theta_i$ for $i=1,\ldots,N^{(s)}$, and $\beta_t^{(s)}$ for $s=1,\ldots,S$, it is straightforward to obtain samples $N^{(s)}_{\alpha}$ of $N_{\alpha}$ and an estimate of the posterior expectation $\widehat{N}_{\alpha}$ by
\be
N^{(s)}_{\alpha} = \sum_{i=1}^{N^{(s)}} \mathbbm{1} \left\{ \varphi^{-1}(\eta_i^{(s)}) >\alpha \right\}, \quad \widehat{N}_{\alpha} = \frac{1}{S} \sum_{s=1}^S N^{(s)}_{\alpha},
\ee
where $\eta^{(s)}_i$ is defined by \eqref{eq:aobsmth}. If a value of $\alpha$ has already been selected then recording samples of $N^{(s)}_{\alpha}$ is all that is necessary to obtain approximate samples from the posterior distribution of $N_{\alpha}$ under the model in \eqref{eq:dpmod}. However, in some cases the heuristics in \S \ref{sec:aobsprops} may not be precise enough to conclusively select a value of $\alpha$. Here we suggest an approach to making a data-driven selection of $\alpha$ \emph{a posteriori}. As a basis for selection of appropriate $\alpha$ values, we introduce the following condition.

\begin{definition}[Informative classes] \label{def:informclass}
Consider a data-augmented likelihood given by $L(x \mid \theta^*, z) = \prod_{i=1}^m L(x_i \mid \theta^*_{z_i})$, where $z_i = 1,\ldots,K$ is a class assignment variable. A class $h$ with associated class-specific parameter $\theta^*_h$ is \emph{informative} if, conditional on the class memberships $z_i, i=1,\ldots,N$, the Fisher information of $\theta^*_h$ is nonzero. Define $\mc H \subset \{1,\ldots,K\}$ as the collection of informative classes. 
\end{definition}

In other words, a latent class is considered informative if the likelihood is not flat in the parameters corresponding specifically to that class. The basic intuition is that if the likelihood is insensitive to the value of some parameters, then those parameters and any functionals of them are not identifiable conditional on the current class assignments. For a Dirichlet process with a base measure $H$ that is a member of the exponential family, it is enough to check whether any \emph{observed} individuals are members of class $h$. Any classes containing at least one observed individual are informative. 

We suggest applying Definition \ref{def:informclass} at each iteration of the Markov chain Monte Carlo algorithm to classify every latent class as informative or uninformative. In particular, since $N_{\alpha}$ is a function of $\theta^*_h$ for every latent class $h$ such that $1-\prod_{t=1}^T\{1-\varphi(\theta^*_h+\beta_t)\} >\alpha$, if any of these latent classes are uninformative then the associated sample for $N_{\alpha}$ may be unreliable. 
Therefore, we propose choosing $\alpha = \alpha_{\inf, q}$, the minimal $\alpha$ such that $N_{\alpha}$ is a function of only informative classes with probability $q \in (0,1)$. This value can be approximated from the quantiles of the sample path $\alpha^{(s)}_{\inf}, s=1,\ldots,S$ where $\alpha^{(s)}_{\inf} = \inf \{\alpha : \theta^{(s)}_h > \alpha \Rightarrow  h \in \mc H \}$. We suggest $q=0.5$ and $q=0.95$ as reasonable default values. Finally, one can compute an analogue of the lower bound estimate $N_{LB}$ by the ergodic average 
\be
N^{(s)}_{LB} = \sum_{i=1}^{N^{(s)}} \mathbbm{1} \left( z_i \in \mc H \right), \quad \widehat{N}_{LB} = \frac{1}{S} \sum_{s=1}^{S} N^{(s)}_{LB}  \label{eq:nlb}
\ee
an estimate of the total number of individuals in informative classes. Like the lower bound estimate of \cite{chao1987estimating}, $N_{LB}$ does not pertain to a concrete population. The empirical properties of these approaches are assessed in Section \ref{sec:sims}.

\section{Simulation studies} \label{sec:sims}
\subsection{Simulation setup} \label{sec:simsetup}
We now conduct a series of simulation studies to evaluate the performance of the Dirichlet process prior on $G^*$ in the $M^{th}$ model in population estimation, estimation of $N_{\alpha}$, and the various approaches to the choice of $\alpha$. Throughout, our approach is compared to the three parametric choices for $G^*$ that are implemented in the \texttt{Rcapture} package for the \texttt{R} environment. All three are exponential tiltings of exponential family distributions, specifically the Poisson, Gamma, and Gaussian distributions. We refer to these as the indirect Poisson, indirect Gamma, and Darroch specifications for $G$, respectively.  These are common parametric choices for $G^*$ in the population estimation literature, primarily for computational convenience, and are described extensively in Supplementary Materials. 

We simulate from the $M^{th}$ model with three choices of $G^*$: (1) the Darroch distribution with $\beta_t = -3.75$ for all $t$ and $\tau = 2$ (see Table \ref{tab:simsumm} in Supplementary Materials); (2) a Normal distribution with mean 2 and variance 14; and, (3) a mixture of two normal distributions with means 0 and -3$\cdot$5, variances 0$\cdot1$ and equal weights on the two components. In each of the simulation scenarios, $N=2000$, $T=4$, and a $\beta_t = \beta$ is set such that $m \in [1000, 1500]$ for most simulated datasets. For each generative model, we generate $R=200$ replicate sets. Analogous results to those that follow are given for seven additional simulation models in the Supplementary Materials. 

\subsection{Estimation of $f^*(\eta)$, $N$, and $N_{\alpha}$}
\begin{figure}[h]
\begin{tabular}{ccc}
\includegraphics[width=0.3\textwidth]{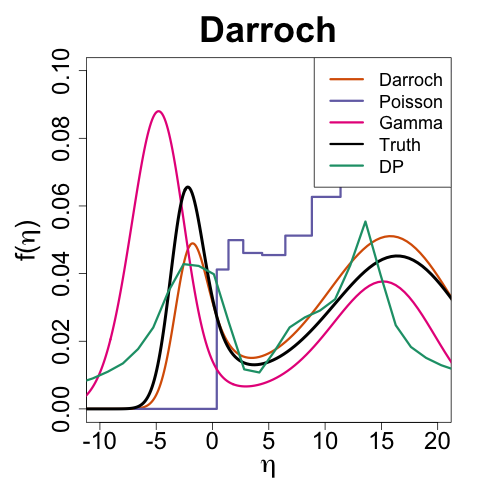} & \includegraphics[width=0.3\textwidth]{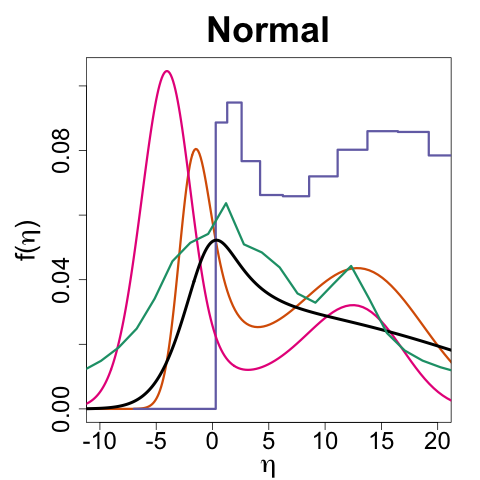} & \includegraphics[width=0.3\textwidth]{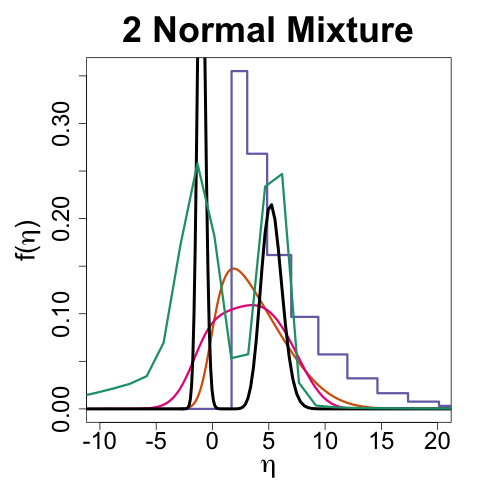}  \\
\end{tabular}
\caption{Estimated density $f^*$ of $\eta$ in the four models considered, plotted with the true $f^*(\eta)$, for three simulation studies. In the figure legend, Dirichlet process is abbreviated DP. } \label{fig:etas}
\end{figure}

We first present results for estimation of $f^*(\eta)$ on the logit scale in each of these models, obtained by transforming from the estimated random effect distribution $\widehat{G^*}$ to the observability distribution $f^*(\eta)$. Figure \ref{fig:etas} shows the estimated density $f^*(\eta)$ obtained using the three parametric choices of $G^*$ and the Dirichlet process model for a single replicate simulation. In general, the Dirichlet process is able to recover the correct shape of $f^*(\eta)$ for $\eta > -5$, which includes $\alpha$-observable individuals for $\alpha>0.01$. The parametric choices performs poorly in this important region in at least one of the simulations. 

We also compare estimates of $N_{\alpha}$ under the Dirichlet process and three parametric estimation models for pre-selected values of $\alpha$ based on the theoretically motivated heuristic in Section \ref{sec:aobsprops}. If we want to achieve probability 0.5 of observing data in any $0.01$ non-negligible set, then by Theorem \ref{thm:hpregion}, $\alpha \approx 0.05$. If we instead want to achieve probability 0.95, then choose $\alpha$ between 0$\cdot$2 and 0$\cdot$3. We calcuate $N_{\alpha}$ for $\alpha \in \{0.05, 0.10, 0.15, 0.20, 0.25\}$, roughly the range of values suggested by the heuristics. 

Table \ref{alphas_tab_reduced} shows root mean square error 
\be
\Delta(\widehat{N}_{\alpha},N_{\alpha}) = \left\{R^{-1} \sum_{r=1}^R \left(\widehat{N}_{\alpha,r} - N_{\alpha,r} \right)^2 \right\}^{1/2}
\ee
where $N_{\alpha,r}$ is the true value of $N_{\alpha}$ for replicate $r$ and $\widehat{N}_{\alpha,r}$ is the associated point estimate. $\Delta$ is computed for each of the four models for $G^*$ and the five $\alpha$ values  and $R=200$ replicates, and results are shown in Table \ref{alphas_tab_reduced}. The table aggregates results over the three simulation scenarios that vary the true $G^*$. For the Dirichlet process model, we estimate the posterior expectation of $N$ and $N_{\alpha}$ by their ergodic averages. For the other models, we use the maximum likelihood estimates of $N$ and $N_{\alpha}$. As predicted by theorem \ref{thm:hpregion}, for every model considered, $\Delta(\widehat{N}_{\alpha},N_{\alpha})$ is considerably smaller for every value of $\alpha$ than $\Delta(\widehat{N},N)$. The best overall value of $\Delta$ across the four estimation models is that for the Dirichlet process model. The Dirichlet process model has a somewhat larger value of $\Delta(\widehat{N},N)$ than the Darroch choice for $G^*$. This is actually a desirable property, indicating a large degree of uncertainty about an essentially unknowable quantity, namely $N_{\alpha}$ for $\alpha = 0$.  

\ifx\dmvdraft\undefined
\begin{table}[ht]
\centering
\begin{tabular}{lccccccc}
& \multicolumn{5}{c}{$\alpha$} & & \\  
Inference model & 0.25 & 0.2 & 0.15 & 0.1 & 0.05 & overall & $\alpha=0$ \\ 
darroch & 97.1 & 244.4 & 296.2 & 307.4 & 312.4 & 264.2 & 317.2 \\ 
  poisson & 144.6 & 300.0 & 359.0 & 377.1 & 396.0 & 328.3 & 444.0 \\ 
  gamma & 110.8 & 234.3 & 264.4 & 251.9 & 262.8 & 232.2 & 1677.7 \\ 
  Dirichlet Process & 120.8 & 211.4 & 193.4 & 123.7 & 201.1 & 174.6 & 796.7 \\ 
\end{tabular}
\caption{Root mean squared error across all simulation scenarios for fixed $\alpha$. Overall aggregates results over the fixed nonzero $\alpha$ values, the $\alpha=0$ column shows results for estimation of $N$.  } 
\label{alphas_tab_reduced}
\end{table}

\fi

Table \ref{coverage} shows coverage of 95 percent intervals for $N$ for each simulation scenario and model for $G^*$. For the three parametric choices of $G^*$, these are 95 percent confidence intervals, and for the Dirichlet process model these are 95 percent posterior credible intervals. Coverage is computed across the 200 replicates of each of the three simulation scenarios. It is clear that when the parametric models are misspecified, they massively and systematically underestimate the uncertainty in $N$, even in cases where there are no modes in the true $G^*$ centered at large negative values. The empirical coverage of the credible intervals for the Dirichlet process model is higher than 95 percent, however, in Bayesian nonparametric models it is not expected that the coverage of credible intervals coincides with the posterior probability of the interval.   

\begin{figure}[h]
\centering
\begin{tabular}{ccc}
\includegraphics[width=0.3\textwidth]{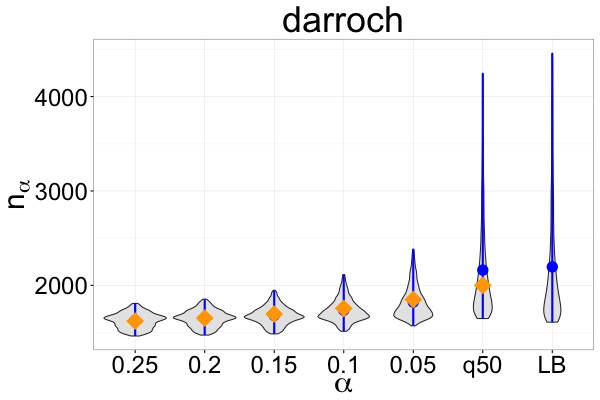} & \includegraphics[width=0.3\textwidth]{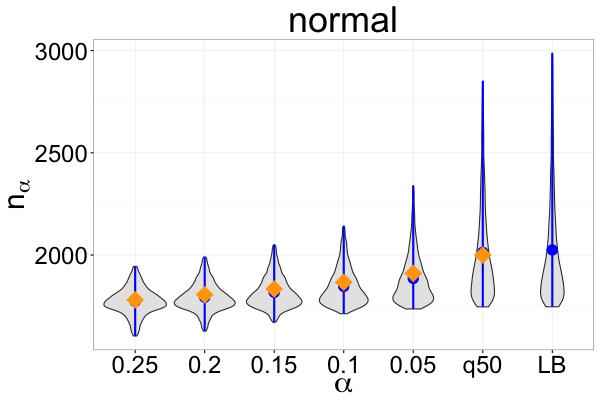}  &  \includegraphics[width=0.3\textwidth]{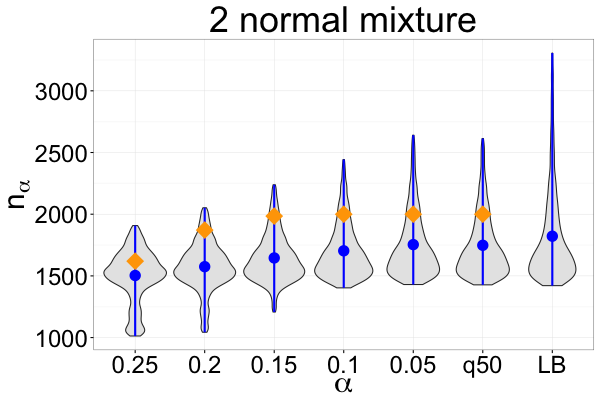} \\
\end{tabular}
\caption{Violin plots showing an estimate of the posterior distribution for $N_{\alpha}$ for several $\alpha$ values for the Dirichlet process model; also shown is the distribution of $N_{\alpha}$ for $\alpha = \alpha_{\inf,0.5}$. These results correspond to a single simulation replicate. The blue dots correspond to posterior means, and the blue lines to a 95 percent credible interval.  } \label{fig:nalphas}
\end{figure}

\ifx\dmvdraft\undefined
\begin{table}[ht]
\centering
\begin{tabular}{rrrrr}
& \multicolumn{4}{c}{inference model} \\
Simulation scenario & Darroch & Poisson & Gamma & Dirichlet process \\ 
darroch & 0.96 & 0.00 & 0.00 & 1.00 \\ 
  normal & 0.70 & 0.00 & 0.00 & 1.00 \\ 
  2 normal mixture & 0.00 & 0.00 & 0.24 & 0.99 \\ 
\end{tabular}
\caption{Coverage of 95\% intervals for total population estimates} 
\label{coverage}
\end{table}

\fi
We next consider the performance of the Dirichlet process model in quantifying uncertainty in $N_{\alpha}$ for the pre-selected values of $\alpha$, $\alpha_{\inf,0.5}$, and $\alpha_{LB}$, the identified population estimate defined in \eqref{eq:nlb}. Figure \ref{fig:nalphas} shows an estimate of the posterior distribution of $N$ and of $N_{\alpha}$ for each considered value of $\alpha$ based on Markov chain Monte Carlo output for a single replicate simulation. Manual examination of results for other simulation replicates was similar. As expected, as $\alpha$ decreases, the posterior distribution of $N_{\alpha}$ becomes more diffuse. However, there is always considerably less uncertainty in $N_{\alpha}$ than in $N$ for every value of $\alpha$ considered, despite the fact that coverage remains at or above 95 percent (see Table \ref{tab:nalpha_cov} in the  Appendix). In many cases, the posterior for $N$ is extremely diffuse and skewed. Thus, the credible intervals reflect the predictions of the theory: estimates of $N$ are in general highly uncertain and unstable. Not only are point estimates of $N_{\alpha}$ more reliable, uncertainty in the value of $N_{\alpha}$ is generally much smaller than uncertainty in the value of $N$. Notably, $\alpha_{\inf,0.5}$ often falls within the range of $\alpha$ values considered and gives similar results. Similar results for six additional simulation scenarios, as well as coverage of 95 percent credible intervals for $N_{\alpha}$ for the Dirichlet process model, are provided in Supplementary Materials. 

\section{Application}
Here we illustrate the method using real data from the now classic snowshoe hare dataset, originally analyzed by \citet{Otis1978} and subsequently re-analyzed by several authors. These data give multiple-recapture history of $n=68$ hares across $T=6$ occasions. Previous analyses, including \citet{cormack:1989:loglinear:cr, Agresti1994, Dorazio2003, Pledger2005}, generally agree that considerable capture heterogeneity across individuals is present. Unsurprisingly, given the small sample size, these analysts have also noted that different models accounting for heterogeneity seem to fit the data comparably well, yet result in different estimates of the population size, $N$. For example \citep{Dorazio2003} reports estimates ranging from $\hat{N} = $76$\cdot$4, with 95\% confidence interval [70$\cdot$2, 86$\cdot$4]), for two-component latent-class $M^{th}$ models; to $\hat{N} =$ 91$\cdot$9, with 95\% confidence interval [74$\cdot$7, 153$\cdot$0], for logistic-normal $M^{th}$ models.

\begin{figure}[h]
\centering
 \includegraphics[width=0.4\textwidth]{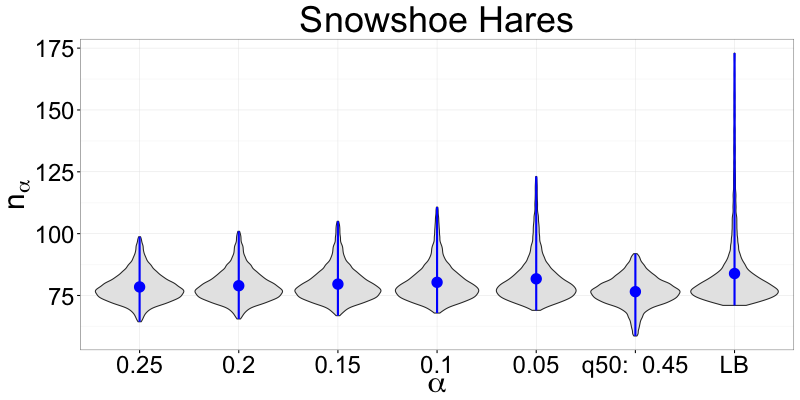}
 \caption{Estimates of the posterior distribution of $N_{\alpha}$ for several fixed $\alpha$ values and $\alpha_{\inf,0.5}$ using the snowshoe hares data. Also shown is the estimated posterior distribution of $N_{LB}$. The blue dots correspond to posterior means, the yellow dots to the true value of $N_{\alpha}$, and the blue lines to a 95 percent credible interval. } \label{fig:hares} 
\end{figure}

Figure~\ref{fig:hares} presents results applying the Dirichlet process $M^{th}$ model. We show estimates of the posterior distribution of $N_{\alpha}$ for fixed values of $\alpha$, for the estimated $\alpha_{\inf,0.5} = 0.45$, as well as for $N_{LB}$. Consistent with the simulation examples, any choice of $\alpha$ results in substantially smaller posterior uncertainty in $N_{\alpha}$ relative to posterior uncertainty in $N$, or even $N_{LB}$. All choices of $\alpha$ result in similar point estimates, around $80$, which are in close agreement with the most conservative previous estimates from heterogeneity models, in the vicinity of $77$. This was expected, as the reduced sample size cannot inform much about individuals with reduced capture probabilities, even if they existed. We also note that, as expected, the uncertainty in $N_{\alpha}$ is larger for smaller values of $\alpha$.

\section{Discussion}
Although it has been clear for some time that population estimation in the presence of substantial capture heterogeneity can be unreliable, the precise nature and scope of this problem has remained obscure. We have clarified the ultimate cause of this problem by equating heterogeneous population estimation to a length-biased density estimation problem. In this context, it is clear that only with strong parametric assumptions about the catchability distribution is it possible to estimate the total population size in finite samples. 

In the absence of a scientific justification for a particular parametric model or strong prior knowledge, we have proposed estimating the ``observable'' population size and have suggested several data-driven approaches to defining this population. The approach results in substantially lower risk estimators with considerably lower uncertainty. Our approach can be viewed as an alternative to using lower bounds on the total population size, and has the advantage that the resulting estimate is more interpretable and automatically converges toward an estimate of the total population size as the sample grows large. The method is also easy to implement with any model for the catchability distribution. As such, the approach provides a default method when selection of a parametric form for the catchability distribution is unwarranted, as is the case in many applications of population estimation. 

There are at least two strong justifications for estimating the observable population size. The first is to view our proposed approach as an extension of the long-standing assumption in population estimation that there are no individuals in the population with zero capture probability. Researchers have long been comfortable with making scientific inferences using traditional population estimates, acknowledging that these estimates exclude all perfectly invisible individuals. Thus, we expect that scientists will find estimates of $N_{\alpha}$, the population size excluding ``nearly invisible'' individuals, similarly useful. Secondly, our results show that estimators of the observable population are biased estimators of $N$ that often have lower risk and are superior in the minimax sense for estimation of the total population size, akin to James-Stein estimators and Bayes estimators. As such, regardless of whether one wishes to assume away individuals that are nearly invisible to the sampling design or not, the approach we suggest has significant advantages over existing methods. 

\section*{Acknowledgments}
This work was inspired by research conducted at Human Rights Data Analysis Group. The authors gratefully acknowledge funding support for this work from the Human Rights Data Analysis Group and the National Institutes of Health. The authors thank David Dunson for helpful comments and suggesting the connection to the length bias literature.

\bibliographystyle{plainnat}
\bibliography{cr_heterogeneity}

\begin{appendix}
\section{Appendix}
\subsection{Proof of Remark \ref{rem:nhatnonhet}}
First consider the case where $p$ is the same for all individuals. Then $m \sim \text{Binomial}(N,p)$, and the log-likelihood is given by
\be
l(m \mid N,p) \propto \log(N!) - \log\{ (N-m)!\} + (N-m) \log(1-p).
\ee
Introducing the penalty factor of $-\frac{1}{2}\log\{ N (N-m)^{-1} \}$, 
\be
\tilde{l}(m \mid N,p) \propto \log\{\Gamma(N+1)\} - \log\{ \Gamma(N-m+1)\} + (N-m) \log(1-p) - \frac{1}{2} \log\{ N (N-m)^{-1} \}.
\ee
This factor is analogous to a continuity correction for the parameter $N$, since, applying Stirling's series
\be
\tilde{l}(m \mid N,p) &= (N+1/2) \log N - N - (N-m+1/2) \log (N-m) + (N-m) \\
&+ \bigO{N^{-1}} + \bigOc{(N-m)^{-1}}  - \frac{1}{2} \log\{ N (N-m)^{-1} \} \\
&= N \log N - N - (N-m) \log (N-m) + (N-m) + \bigO{N^{-1}} + \bigOc{(N-m)^{-1}},
\ee
so the penalty term gives an asymptotically exact expression for the log-likelihood of $\textnormal{Binomial}(N-1/2,p)$. Treating $N$ as a continuous parameter, differentiating and setting to zero
\be
\frac{\partial }{\partial N} \tilde{l}(m \mid N,p) = \psi(N+1) - \psi(N+1-m) + \frac{m}{2N(N-m)} + \log(1-p) &= 0
\ee

which implies
\be
\exp[ \psi(N+1) - \psi(N+1-m) + m \{2 N (N-m)^{-1} \} ] &= \frac{1}{1-p},
\ee
where $\psi$ is the digamma function. From \cite{qi2011inequality}, for $x \in (0,\infty)$,
\be
\log(x) + \frac{1}{2x} - \frac{1}{12 x^2} < \psi(x+1) < \log(x) + \frac{1}{2x},
\ee
so $\psi(x+1) = \log(x) + (2x)^{-1} - h(x)$ for a function $(12 x^2)^{-1} < h(x) < 0$. Substituting
\be
\frac{N}{N-m}  \exp\{ (2N)^{-1} - \{2(N-m)\}^{-1} + m \{2 N (N-m)^{-1} \} - \{h(N)-h(N-m)\}  \} &= \frac{1}{1-p} \\
\frac{N}{N-m}  \exp\left[ -m\{2 N (N-m)\}^{-1} + m \{2 N (N-m)^{-1} \}  - \{h(N)-h(N-m)\} \right] &= \frac{1}{1-p} \\
\frac{N}{N-m}  \left( 1 - \{h(N)-h(N-m)\} + \bigOs{\{h(N)-h(N-m)\}^2}       \right) &= \frac{1}{1-p} \\
\frac{N}{N-m}  \left( 1 - \bigO{N^{-2}}  \right) &= \frac{1}{1-p}
\ee
Solving
\be
N \left[ 1 - \bigO{N^{-2}}  \right]   &= \frac{N-m}{1-p} \\
N \left( 1- \frac{1}{1-p} \right)  - \bigO{N^{-1}}  &= \frac{-m}{1-p} \\
Np + \bigO{N^{-1}} &= m \\
\widehat{N} &= \frac{m}{p} - \bigO{m^{-1}}.
\ee

This immediately gives a result for $p \sim F$ where $F$ consists of $J<\infty$ atoms. Suppose $\PR{P=p_j} = w_j$, and let $m_1,\ldots,m_J$ be the counts of observed units with $p=p_j$. Then
\be
(N-m,m_1,\ldots,m_J) \sim \textnormal{Multinomial}\left\{ N,\left( 1-\sum_j w_j p_j, w_1 p_1,\ldots,w_J p_J \right) \right\},
\ee
so
\be
m \sim \textnormal{Binomial}\left( N,\sum_j w_j p_j \right),
\ee
giving
\be
\widehat{N} &= \frac{m}{E_F(P)} - \bigO{m^{-1}}.
\ee

\subsection{Proof of Theorem \ref{thm:hpregion}}
Let $Y$ be the random variable defined by 
$$Y = \bone{p \in A} \bone{\sum_{t=1}^T X_{t}>0},$$ 
where $X_t$ is the random variable of which $x_{it}$ is an observation. Clearly, $Y$ is binary, and the event $\{Y = 1\}$ occurs exactly when $p \in A$ and $S = \sum_{t=1}^T X_t > 0$. Since $\PR{S>0} = p$ and $A$ is $\alpha$-observable, $\PR{S > 0 \mid p \in A} \ge \alpha$. Since $A$ is $f,\epsilon$ non-negligible, it follows that $\PR{Y > 0} \ge \epsilon \alpha$.

Therefore, $M(A)$ is the sum of independent binary random variables $Y_1,\ldots,Y_N$ each having $\PR{Y_i = 1} \ge \epsilon \alpha$. So, $M(A)$ stochastically dominates a $\Binom{N}{\alpha \epsilon}$ random variable. So then
\be
\PR{M(A) > 0} \ge 1-(1-\alpha \epsilon)^N
\ee

\subsection{Proof of Theorem \ref{thm:empmeasure}}
We have $\widehat{N} = \sum_{i=1}^m p_i^{-1}$. We will need
\be
E_G(1/P) &= \int p^{-1} G(dp) = \int p^{-1} p F(dp) \{E_F(P)\}^{-1}  = \{E_F(P)\}^{-1} \\
E_G(P^{-2}) &= \int p^{-2} G(dp) = \int p^{-2} p F(dp) \{E_F(P)\}^{-1}  = E_F(P^{-1}) \{E_F(P)\}^{-1} 
\ee
so
\be
\text{var}_G(P^{-1}) &= E_F(P^{-1}) \{E_F(P)\}^{-1} - \{E_F(P)\}^{-2}.
\ee

First compute the variance
\be
\Var{\widehat{N}} &= \Varc{E(\widehat{N} \mid m)} + \EXc{\Var{\widehat{N} \mid m}} \\
&=  \Varc{m E_G(P^{-1}) } + \EXc{m\, \text{var}_G(P^{-1})} = \{E_F(P^{-1})\}^2 \Var{m} + \EX{m} \text{var}_G(P^{-1}) \\
&= N \{E_F(P)\}^{-2} [E_F(P) - \{E_F(P)\}^2] + N E_F(P) [E_F(P^{-1}) \{E_F(P)\}^{-1} - \{E_F(P)\}^{-2}] \\
&= N [ \{E_F(P)\}^{-1} - 1] + N [E_F(P^{-1}) - \{E_F(P)\}^{-1}] \\
&= N \{ E_F(P^{-1}) - 1 \},
\ee
where the third step used the variance of $m$ derived in the proof of Theorem \ref{thm:histmse}. Since
\be
E(\widehat{N}) &= E\{m E_G(P^{-1})\} = N E_F(P) \{E_F(P)\}^{-1},
\ee
the estimator is unbiased, so
\be
\Delta^2(\widehat{N},N) = N \{ E_F(P^{-1}) - 1 \}.
\ee

\subsection{Proof of Theorem \ref{thm:histmse}}
For this result we have the assumptions that $F$ has density $f$ that is twice continuously differentiable, which implies that $G$ also has a density. Consider a histogram estimator of a density $g$ on the unit interval with bin width $h$, which is chosen such that $1/h$ is an integer. Using \eqref{eq:fhat}, the estimator of $N$ conditional on $m$ becomes
\be
\widehat{N} &= m \left\{ \int p \widehat{f}(p) dp \right\}^{-1} = m \left\{ \int p \frac{p^{-1} \widehat{g}(p)}{\int_0^1 p^{-1} \widehat{g}(p) dp } dp \right\}^{-1} \\
&= m \left\{ \frac{ \int  \widehat{g}(p) dp}{\int_0^1 y^{-1} \widehat{g}(p) dp }  \right\}^{-1} = m \left\{ \frac{1}{\int_0^1 p^{-1} \widehat{g}(p) dp }  \right\}^{-1} = m \int_0^1 p^{-1}  \widehat{g}(p) dp.  
\ee
Let $B_j$ be the count in the $j$th bin. For $\widehat{g}$ the histogram estimator we have
\be
\Var{\widehat{N} \mid m} &= m^2 \Vars{\int p^{-1}  \widehat{g}(p) dp} \\
&= m^2 \Vars{ \sum_{j=0}^{1/h-1} \int_{hj}^{h(j+1)} y^{-1} \widehat{g}(p) dp } = \Vars{ \sum_{j=0}^{1/h-1} h p_j^{-1} \widehat{g}(p_j) } \label{eq:mvtapp} \\
&= m^2 \Vars{ \sum_{j=0}^{1/h-1} h p_j^{-1}  m^{-1} h^{-1} B_j } = \Vars{ \sum_{j=0}^{1/h-1} p_j^{-1} B_j }  \\
&= \sum_{j=0}^{1/h-1}  p_j^{-2} \Var{B_j} + 2 \sum_{j=0}^{1/h-1} \sum_{k=j+1}^{1/h-1} p_{j}^{-1} p_{k}^{-1} \text{cov}(B_{j},B_{k}).
\ee
where $p_j \in [hj,hj+h]$ and we applied the mean value theorem in \eqref{eq:mvtapp}. The $B_j$ are jointly $\text{Multinomial}[m,\{G(h)-G(0),G(2h)-G(h),\ldots, G(1)-G(1-1/h)\}]$, where $G(t) = \int_0^t g(p) dp$, so
\be
\Var{\widehat{N} \mid m} &= \sum_{j=0}^{1/h-1}   p_j^{-2} m \{G(hj+h)-G(hj)\} \{1-G(hj+h)+G(hj)\} \\
&-  2 \sum_{j=0}^{1/h-1} \sum_{k=j+1}^{1/h-1} p_{j}^{-1} p_{k}^{-1}  m \{G(hj+h)-G(hj)\}\{G(hk+h)-G(hk) \} \\
&= \sum_{j=0}^{1/h-1}  p_j^{-2} m h g(p_j^*) \{1- h g(p_j^*)\} -  2 \sum_{j=0}^{1/h-1} \sum_{k=j+1}^{1/h-1} p_{j}^{-1} p_k^{-1} m h^2 g(p^*_j) g(p^*_k) \\
&= m \int_0^1  p^{-2} \{g(p) - h g^2(p)\} dp -  2m \int_{p_1=0}^1 \int_{p_2=p_1}^1 p_1^{-1} p_2^{-1} g(p_1) g(p_2) dp_1 dp_2 + \bigO{mh},
\ee
where again we applied the mean value theorem, and $p_j^* \in [hj,hj+h]$. Using \eqref{eq:obsdens} 
\be
\Var{\widehat{N} \mid m} &= m \int_0^1   p^{-2} \left[\frac{p f(p)}{\int_0^1 p f(p) dp} - h \frac{p^2 f^2(p)}{\left\{\int_0^1 p f(p) dp \right\}^2}  \right] dp \\
&-  2m \int_{p_1=0}^1 \int_{p_2=p_1}^1 p_1^{-1} p_2^{-1} \frac{p_1 f(p_1)}{\int_0^1 p f(p) dp} \frac{p_2 f(p_2)}{\int_0^1 p f(p) dp} dp_1 dp_2 +  \bigO{mh} \\
&= m  \left[ \frac{E_f(P^{-1}) }{E_f(P)} - h \frac{E_f\{f(P)\}}{\{E_f(P)\}^2} - \frac{1}{\{E_f(P)\}^2}  \right]  + \bigO{mh} \label{eq:varsimp} \\
&= m  \left[ \frac{E_f(P^{-1}) }{E_f(P)} - \{E_f(P)\}^{-2}  \right]  + \bigO{mh}
\ee

In \eqref{eq:varsimp}, we used the fact that
\be
\int_{p_1=0}^1 \int_{p_2=p_1}^1 f(p_1) f(p_2) dp_1 dp_2 
\ee
is the integral of the product density $f(p_1,p_2) = f(p_1) f(p_2)$ over the region of the unit square above the line $p_1=p_2$. Since $f(p_1,p_2)$ is symmetric about the line $p_1=p_2$, we have
\be
\int_{p_1=0}^1 \int_{p_2=p_1}^1 f(p_1) f(p_2) dp_1 dp_2 = \frac{1}{2} \int_{p_1=0}^1 \int_{p_2=0}^1 f(p_1) f(p_2) dp_1 dp_2 = \frac{1}{2}.
\ee
Since $E(m \mid N, f) = N \{E_f(P)\}$, we have
\be
\Var{\widehat{N} \mid N,f} &= N  \left[ E_f(P^{-1}) - \{E_f (P)\}^{-1}  \right]  + \bigO{Nh}.
\ee

Now we compute the absolute bias
\be
E(\widehat{N} \mid m, \widehat{g}) - E(N \mid m,g) &= m E \left[ \int_{0}^1 p^{-1} \{ \widehat{g}(p) - g(p)\} dp \right] \\
&= m \int_{0}^1p^{-1} E\{ \widehat{g}(p) - g(p)\} dp  = m \int_{0}^1 p^{-1} \text{Bias}\{ \widehat{g}(p) \mid m\} dp.
\ee
The bias of $\widehat{g}(p)$ is $g'(p)(hj+h/2-p) + \bigO{h^2}$, see \cite{wasserman2006all}, so we have
\be
\left| E(\widehat{N} \mid m, \widehat{g}) - E(N \mid m,g) \right| &= m \left| \sum_{j=0}^{1/h-1} \int_{hj}^{hj+h} p^{-1} \left[g'(p)(hj+h/2-p) + \bigO{h^2} \right]  dp \right| \\
&= m \left| \sum_{j=0}^{1/h-1} h p_j^{-1} \left[ g'(p_j)(hj+h/2-p_j) + \bigO{h^2} \right] \right| \\
&\le m \sum_{j=0}^{1/h-1} h p_j^{-1} \left[ |g'(p_j)| |hj+h/2-p_j| + \bigO{h^2} \right]  \\
&\le \frac{m h}{2} \left( \int_0^1 p^{-1} |g'(p)| dp + \bigO{h} \right) +  \bigO{mh^2}  \\
&\le \frac{m h}{2 \int_{0}^1 p f(p) dp} \int p^{-1}  |\{p f(p)\}'|  dp + \bigO{mh^2}   \\
&\le \frac{m h}{2 \int_0^1 p f(p) dp} \int p^{-1} f(p) + |f'(p)| dp + \bigO{mh^2}.
\ee
where we applied the mean value theorem in the second step. So 
\be
\left\{ \text{Bias}(\widehat{N} \mid m) \right\}^2 &\le \left(\frac{m h}{2 E_f(P)} \left[ E_f(P^{-1}) + E_f\{ |f'(P)|\}  \right] \right)^2 + \bigO{m^2 h^3}.
\ee
Now we just need to calculate $E(m^2 \mid N, f)$. Let $x_i \sim \textnormal{Bernoulli}\{p_i\}$ be the indicator that the $i$th unit is observed, then
\be
\Var{m \mid N,f} &= \Varc{E(m \mid p_1,\ldots,p_N)} + E\{ \Var{m \mid p_1,\ldots,p_N} \} \\
&= \Varc{\sum_{i=1}^N p_i } + E \left\{ \Var{ \sum_{i=1}^N x_i } \right\} \\
&= \sum_{i=1}^N \Var{p_i}  + E \left\{ \sum_{i=1}^N p_i(1-p_i)  \right\} \\
&= \sum_{i=1}^N E(P_i^2)-\{E(P_i)\}^2 + \sum_{i=1}^N E\{P_i(1-P_i)\} = N [E_f(P) - \{E_f(P) \}^2],
\ee
so 
\be
E(m^2 \mid N,f) &= N [E_f(P) - \{E_f(P) \}^2] + N^2  \{E_f(P) \}^2 \\
&= N E_f(P) + (N^2-N) \{E_f(P) \}^2.
\ee
So finally
\be 
\bigO{Nh} \le \Delta^2(\widehat{N},N) - N   \left[ E_f(P^{-1}) - \{E_f(P)\}^{-1} \right]  \le \frac{N^2 h^2}{4} \left\{ E_f(P^{-1}) + E_f\{|f'(P)|\} \right\}^2 + \bigO{Nh},
\ee
which gives the result.

 \FloatBarrier
\section{Supplementary Material}
\subsection{Table of simulation scenarios}
Table \ref{tab:simsumm} shows the choices of $G^*$ for each of the ten simulation studies that contribute to the root mean square error results in Table \ref{alphas_tab_reduced} in the main text; the first three are those discussed in the main text. 

\begin{table}
\centering
\setlength{\extrarowheight}{0.3cm}
\begin{tabular}{lll}
Scenario & Model & $G^*(d\theta) \propto$  \\
1 & Darroch & $\sum_{t=1}^T w_t \text{No}(\tau^2 t, \tau^2)$   \\
2 & Normal & $\text{No}(0,\tau^2)$  \\
3 & two normal mixture & $\sum_{h=1}^2 w_h  \text{No}(\mu_h,\tau^2_h)$ \\
4 & truncated normal & $\text{No}(0,\tau^2) \bone{\theta \in [a,b]}$ \\
5 & indirect gamma & $\sum_{t=1}^T w_t \text{Gam}\{\tau, (\lambda+t)^{-1} \}$  \\
6 & atoms & $ \sum_{h=1}^K w_h \delta_{\theta^*_h}(d\theta)$ \\
7 & multi-normal mixture & $\sum_{h=1}^K w_h  \text{No}(\mu_h,\tau^2_h)$ \\
8 & multi-t mixture & $\sum_{h=1}^K w_h  \text{t}_{\nu_h}(\mu_h,\tau^2_h)$ \\
9 & Normal, small variance & $\text{No}(0,\tau^2)$ \\
10 & Normal, $T=7$ & $\text{No}(0,\tau^2)$ \\
\end{tabular}
\caption{Data generating models for simulations. The weights $w_t$ for the Darroch and indirect Gamma models are given in \cite{rivest2007applications}.  } \label{tab:simsumm}
\end{table}

\begin{table}
\centering
\setlength{\extrarowheight}{0.3cm}
\begin{tabular}{lll}
Scenario & parameters & $\beta$  \\
1 & $\tau^2=2$ & -3$\cdot$75 \\
2 & $\tau^2=14$ & 2  \\
3 & $w_h=$ 0$\cdot$5, $\tau^2_h=$ 0$\cdot$1, $\mu_{1:2}=$ (0,-3$\cdot$5) & 1  \\
4 & $\tau^2=12$, $a=-2$, $b=\infty$ & -1 \\
5 & $\tau=1$, $\lambda=$ 3$\cdot$5 & -2  \\
6 & $w \sim \text{Dirichlet}(1,\ldots,1)$, $\theta^* = (-5,$ -3$\cdot$7, -3$\cdot$2, -2$\cdot$75, 0, 1, 3) & 0 \\
7 & $w \sim \text{Dirichlet}(1,\ldots,1)$, $\mu=$(-4,-2,-2$\cdot$5,-2$\cdot$25,0,1,3), $\tau^2=$(0$\cdot$1,2,0$\cdot$05, 0$\cdot$1, 1, 0$\cdot$1, 6)  & 1 \\
8 & $\nu=3$, $w,\mu,\tau^2$ same as Scenario 7 & 1 \\
9 & $\tau^2=$ 0$\cdot$1 & 0 \\
10 & $\tau^2=10$ & -1 \\
\end{tabular}
\caption{Parameters for data generating models used in simulations. } \label{tab:simparam}
\end{table}

\subsection{Parametric choices for $G^*$ in the $M^{th}$ model} \label{sec:mthparam}
In this section, we review common parametric choices for $G^*$ in the $M^{th}$ model. Our purpose is (1) to provide background that allows comparison of the method we propose in \S \ref{sec:dp} to common alternatives; (2) to show that while making a parametric choice for $G^*$ ensures that $N$ is identifiable, when the model is misspecified, the results can be highly misleading, and (3) to show that the while the most common choices for $G^*$, which are exponentially tilted exponential family distributions, are mathematically convenient, they are both highly inflexible and imply very strong assumptions about $G^*$ in $\alpha$-unobservable regions.

Certain parametric choices for $G^*$ lead to convenient analytical expressions for the cell probabilities $\pi_{x}$ in $M^{th}$ models. For this reason, much of the literature has focused on distributions $G^*$ indirectly defined so that $G^*_0(\theta) \sim G^*(\theta \mid x = \zeta)$, for $G^*_0(\theta)$ in a common parametric family such as the normal. Doing so ensures that, after integrating out the random effects, the likelihood belongs to the exponential family. Moreover, the expected cell counts marginal of the individual-level random effects can be expressed in closed form as
\be
E_{\theta}\{n(x)\} = \exp \left \{\beta_0 + \sum_t x_t \beta_t  + \phi_{G^*_0}\left(\sum_t x_t\right) \right \}, \label{eq:cellcounts}
\ee
\noindent where $\phi_{G^*_0}(j)$ is the cumulant generating function of $G^*_0(\theta)$, i.e. $\phi_{G^*_0}(j) = \log \int_{-\infty}^{\infty} e^{j\theta} G^*_0(\theta)$. 
Thus, if the cumulant generating function of $G^*_0(\theta)$ can be easily calculated, the expected cell counts and thus the cell probabilities marginal of $\theta$ are also easily computed. 
Table~\ref{tab:dists} shows three distributions commonly used in capture-recapture.

\begin{table}
\centering
\caption{\label{tab:dists} Common distributions used for $G^*_0(\theta)$.}
\begin{tabular}{ l | l } \hline 
Distribution Name &  Definition  \\ \hline 
Darroch & $g^*_0(\theta) = N( 0,\tau)$\\
indirect Poisson & $\theta = \log(2)X$ \hspace{3mm} $X \sim \text{Poisson}(\tau)$\\ 
indirect Gamma & $\theta = -X$ \hspace{3mm} $X \sim \text{Gamma}(3.5, \tau)$
\end{tabular}
\end{table}

 
Mixing distributions defined in this way can be problematic.  As \citet{baillargeon2007rcapture} show, whenever $G^*_0(\theta)$ is in the exponential family, the resulting $G^*(\theta)$ will be a mixture of exponentially tilted exponential family distributions. Therefore, in general $G^*(\theta)$ will be a multi-modal distribution governed entirely by the parameters of $G^*_0(\theta)$. Moreover, if $G^*_0(\theta) = G^*_0(\theta; \tau)$ has only one parameter, as is typically the case, $\tau$ will define the shape and location of all the modes. For example, if $G^*_0 \sim N(0,\tau)$ -- the Darroch distribution in Table~\ref{tab:dists} -- then $G^*(\theta)$ will be a mixture of $T+1$ Gaussian distributions, with means $t\tau$ and variance $\tau$ for $t=0, 1, \ldots, T$. Thus, the variance $\tau$ in $G^*_0(\theta)$ defines not only the variance of each of the mixture components but also their locations. 

In practice, the consequence of choosing these inflexible parametric distributions is that in order to fit the shape of $G^*$ in the observed region, a mode may be introduced in the unobserved region. This can result in highly misleading results, as there is essentially no data to inform about the shape of the distribution in the unobserved region. 

To illustrate this point, we simulate data from the model in (\ref{eq:mod}) using different choices for $G^*$, then estimate $G^*$ assuming that it is a member of each of the parametric classes in Table \ref{tab:dists}. We choose three different true distributions for $G^*$: Darroch, Gaussian ($\theta \sim N(0,\sigma^2)$), and a mixture of two normals. In the first case, the model is exactly Darroch and the other two parametric choices for $G^*$ are misspecified. In the other two data generation cases, all three of the models used for estimation are misspecified.   

\begin{figure}[h]
\centering
\begin{tabular}{ccc}
\includegraphics[width = 0.3\textwidth]{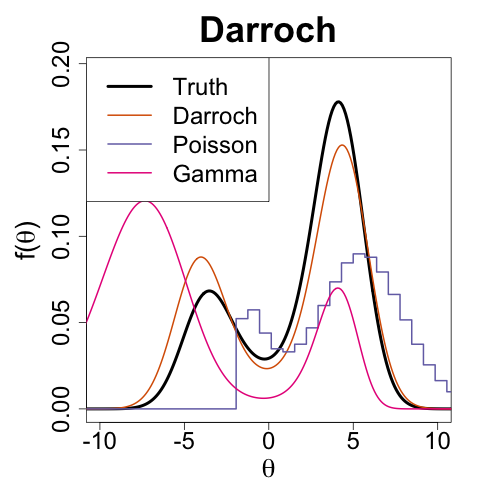} &
\includegraphics[width = 0.3\textwidth]{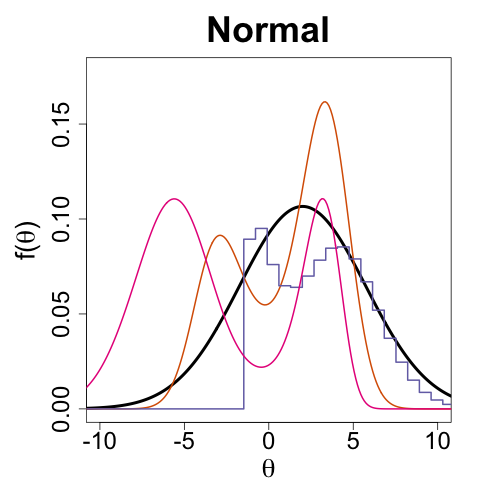} &
\includegraphics[width = 0.3\textwidth]{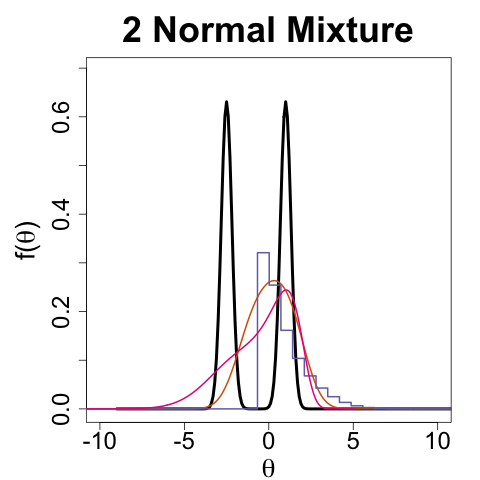} \\
\end{tabular}
\caption{\label{fig:examples} Estimates of the random effects distribution under three estimation models. }
\end{figure}

Figure \ref{fig:examples} shows results. In each panel, the true density $g^*$ is shown in black, and the estimated density $\widehat{g^*}$ under each of the parametric models is displayed on the same plot; we refer to these estimates as $\widehat{g^*}_{D}$ for the Darroch estimate, $\widehat{g^*}_{G}$ for the indirect Gamma estimate, and $\widehat{g^*}_{P}$ for the indirect Poisson estimate. These estimates were obtained as transformations of the output of the \texttt{Rcapture} package $M^{th}$ model fit to simulated data from a population of size $N=2000$. As expected, in the first simulation, $\widehat{g^*}_D$ recovers $g^*$ almost exactly, while $\widehat{g^*}_G$ and $\widehat{g^*}_P$ perform poorly; $\widehat{g^*}_G$, in particular, contains a large mode at approximately $\theta = -7$ that is not present in the data generating model. A unit with $\theta = -7$ has a 0.36\% chance of being observed on any of the four lists, making these estimated individuals $0.01$-unobservable. Even if all 2000 individuals in the population had $\theta \le -7$, we would expect to observe only seven of them. Similar behavior of $\widehat{g^*}_G$ is evident when the data generating model is Normal. In this case, $\widehat{g^*}_D$ also introduces modes where none exist in the true $g^*$, though they do not reside as far in the left tail as those in $\widehat{g^*}_G$. In both the first and second simulations, $\widehat{g^*}_P$ is improperly truncated on the left. Interestingly, in the third simulation, there is a substantial mode around $\theta = -3$ that is completely missed by all three estimators. 

\begin{figure}[h]
\centering
\begin{tabular}{ccc}
\includegraphics[width = 0.3\textwidth]{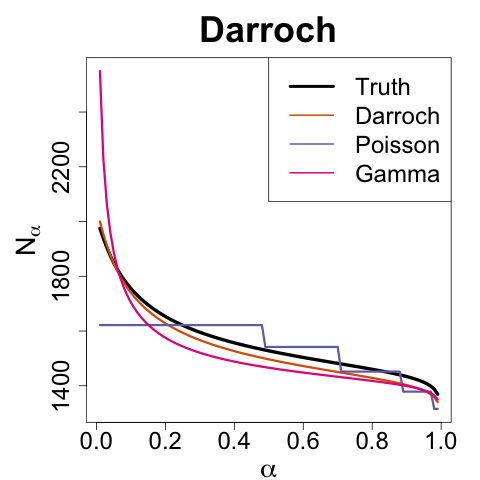} &
\includegraphics[width = 0.3\textwidth]{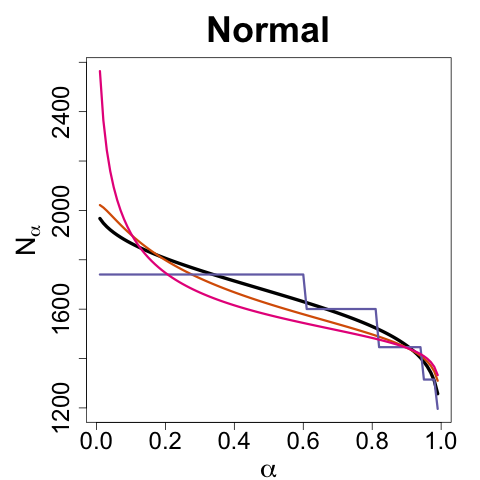} &
\includegraphics[width = 0.3\textwidth]{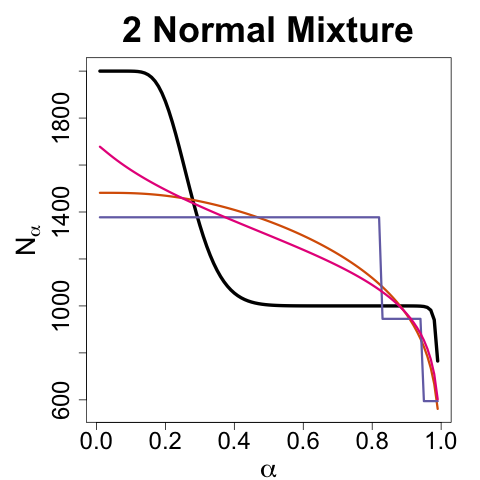} \\
\end{tabular}
\caption{\label{fig:alphaparam} Comparison of expectation of $N_{\alpha}$ under true model to estimates using the three parametric models considered. }
\end{figure}

Although the three parametric distributions considered are generally ill-suited to characterizing $g^*$ in these simulations, the strategy of estimates of $N_{\alpha}$ can nonetheless be accurate in many cases. Figure \ref{fig:alphaparam} compares the expectation of the $\alpha$-observable population size in the true data generating model to the point estimates of $N_{\alpha}$ using $\widehat{g^*}_D, \widehat{g^*}_G,$ and $\widehat{g^*}_P$. In the first two simulations, estimates of $N_{\alpha}$ using either $\widehat{g^*}_D$ or $\widehat{g^*}_G$ are fairly accurate for values of $\alpha$ as small as 0.1. The third simulation is the most challenging, with none of the estimators producing reliable estimates over a consistent range of $\alpha$ values. This reflects that the shape of the true $g^*$ even in the observable region is simply not represented by any members of the three parametric classes; use of a more flexible model would likely ameliorate the problem considerably.

\subsection{MCMC algorithm for the nonparametric $M^{th}$ model}
We use a truncated stick-breaking prior approximation of the Dirichlet process, see \cite{ishwaran2011gibbs}. Let $K$ be the stick-breaking truncation level. The complete data model can be represented as 
\begin{align*}
x_{it} &= I(\gamma_{ij} > 0) \text{ for $i = 1,...,N$ and $t = 1,...,T$}\\
\gamma_{it} &\sim N(\theta_{z_i} + \beta_t, 1)\\
\beta_t &\iid N(\phi, \sigma^2) \text{ for $t = 1,..., T$}\\
z_i &\iid \discreted(\{1,...,K\}, (\pi_1,...,\pi_K)) \text{ for $i=1,...,N$}\\ 
\theta^*_h &\iid N(\mu, \tau^2)  \text{ for } h=1,...,K\\
\nu_h &= \nu^*_h \prod_{l < h} (1 - \nu^*_l) \text{ for } h=1,...,K\\
\nu^*_h &\iid \betad(1,\alpha_0), \text{ for } h < K \text{ and } \nu^*_K = 1\\
\alpha_0 &\sim \gammad(a, b), \quad p(N) \propto 1/N
\end{align*}

Let $n$ be the number of vectors such that $\bx_i \neq (0,...,0)$. We re-label all the variables that depend on the index $i = 1,...,N$ so that $\bx=(0,...,0)$ for $i > m$. Let $\omega = (\omega_1,...,\omega_K)$ where $\omega_h = \sum_{i=m+1}^N I(z_i = h)$ is the count of non-observed units in latent class $h$. Similarly, let $m_h = \sum_{i=1}^m I(z_i = h)$ be the number of observed units in latent class $h$. Finally define $\Gamma^0_{th} = \sum_{i: i>m, z_i = h} \gamma_{it}$.


\begin{enumerate}
	\item \textbf{Sample from $(z_i)_{i \leq m}$}: For $i=1,...,m$, sample $z_i \sim \discreted(\{1,...,K\},(q_1,...,q_{K})),$ 	with $q_h \propto \nu_{h} \exp\left(-T(\theta^*_h)^2/2 - \theta^*_h \left(\sum_{t=1}^T \beta_{t} - \sum_{t=1}^T \gamma_{it}\right)\right)$. 
    
    \item \textbf{Sample from $(\gamma_{it})_{i\leq m}$}: For $i=1,...,m$\
    \begin{align*}
    \gamma_{it} \sim \theta^*_{z_i} + \beta_t + \begin{cases}
    \text{TNorm}(0,1,[-\theta^*_{z_i} - \beta_t, +\infty)) & \text{ if } x_{it} = 1\\
    \text{TNorm}(0,1,(-\infty, -\theta^*_{z_i} - \beta_t)) & \text{ if } x_{it} = 0
    \end{cases}
    \end{align*}

	\item \textbf{Joint sampling $(\beta, \theta^*)$}: Make $\nu_h = m_h + \omega_h$ for $h=1,...,K$, and sample $(\beta, \theta^*)' \sim N(\mu, \Sigma)$ for 
	\begin{align*}
	\Sigma = \left[
	\begin{array}{c|c}
	(1/\sigma^2 + N) I_T & \left( 
    	\begin{array}{cccc}
         \nu_1 & \nu_2 &  \dots \nu_K\\
         \nu_1 & \nu_2 &  \dots \nu_K\\
         \dots\\
         \nu_1 & \nu_2 &  \dots \nu_K
    	\end{array}\right)\\
    \hline
    \left(\begin{array}{cccc}
    \nu_1 & \nu_1 & \dots & \nu_1\\
    \nu_2 & \nu_2 & \dots & \nu_2\\
    \dots\\
    \nu_K & \nu_K & \dots & \nu_K\\
    \end{array} \right)
    & 1/\tau^2 I_K + T \left(
        \begin{array}{cccc}
            \nu_1 & 0 & \dots & 0\\
            0 & \nu_2  & \dots & 0\\
            \vdots &\vdots &\ddots&\vdots\\
            0 & 0 & \dots &\nu_K\\
        \end{array}
        \right)
	\end{array}
	\right]^{-1}
	\end{align*}
and 
$$
\mu = \Sigma
\left[
\begin{array}{c}
\phi/\sigma^2  + \sum_{i=1}^m \gamma_{it} + \sum_{h=1}^K \Gamma^0_{1h}\\
\phi/\sigma^2  + \sum_{i=1}^m \gamma_{it} + \sum_{h=1}^K \Gamma^0_{2h}\\
\vdots\\
\phi/\sigma^2  + \sum_{i=1}^m \gamma_{it} + \sum_{h=1}^K \Gamma^0_{Th}\\
\hline
1/\tau^2  + \sum_{i:i\leq m, z_i=1} \gamma_{it} + \sum_{t=1}^T \Gamma^0_{t1}\\
1/\tau^2  + \sum_{i:i\leq m, z_i=2} \gamma_{it} + \sum_{t=1}^T \Gamma^0_{t2}\\
\vdots\\
1/\tau^2  + \sum_{i:i\leq m, z_i=K} \gamma_{it} + \sum_{t=1}^T \Gamma^0_{tK}\\
\end{array}
\right]
$$

	\item \textbf{Sample from $\nu$}: For $h=1,...,K-1$ sample 
	$$\nu^*_h \sim \betad\left(1+ \nu_h, \alpha_0 + \sum_{l=h+1}^{K} \nu_l \right)$$ 
	where $\nu_h = m_h + \omega_h$. Let $\nu^*_K=1$ and make $\nu_h = \nu^*_h\prod_{l<h}(1-\nu^*_l)$ for all 
	$h=1,...,K$.
	\item \textbf{Sample from} $\alpha_0 \sim \text{Gamma}(a - 1 + K, b - \log\nu_{K})$
	
	\item \textbf{Sample from $(N,(z_i)_{i>m}, (\gamma_{it})_{i>m})$}: For this we use a conditional sampling strategy, similar to \cite{Manrique:LCM:2016:Capture:Recapture}. We factor this joint distribution into:
	$$
	p(N,(z_i)_{i>m}, (\gamma_{it})_{i>m} |\theta^*, \beta, \nu) = p(N|\theta^*, \beta,\nu^*)p((z_i)_{i>m}|\theta^*, \beta, N,\nu) p((\gamma_{it})_{i>m}|\theta^*,\beta,N, (z_i)_{i>m},\nu)
	$$
	and sample from the three partial conditional distributions sequentially. Since we only need $N$, $\boldsymbol\omega$, $\Gamma^0_{jk}$, we sample these variables instead.
	
	\begin{enumerate}
	\item Sample from $N|\theta, \beta,\pi$: Let $N_0 = N - n$. The partial conditional distribution of $N_0$ for prior $p(N) \propto 1/N$
	$$
    N_0 \sim \text{NegBinomial} \left(m, 1 - \sum_{h=1}^{K} \rho_h \right)
	$$
	where $\rho_h = \nu_h \prod_{t=1}^T \Phi(-\theta^*_h - \beta_t)$ and $\Phi(\cdot)$ is the distribution function of a standard normal distribution. Set $N = N_0 + m$.
	
	\item Sample from $\boldsymbol\omega \sim \text{Multinomial}(N_0, (q_1,...,q_K))$
    where $q_k \propto \rho_k$.
    \item Sample $\Gamma^0_{th}$: For $i=1,...,\omega_h$ get $\gamma^0_{ith} \iid \text{TNorm}(\theta^*_{k} +\beta_t, 1, [0, \infty))$ and make $\Gamma^0_{th} = \sum_{i=1}^{\omega_h} \gamma^0_{ith}$
	\end{enumerate}
\end{enumerate}

\subsection{Additional information on the $M^{th}$ model} \label{sec:mthadd}
If we choose $\varphi = \ell$, the likelihood simplifies to $\prod_{t=1}^T \frac{e^{(\theta_i+\beta_t)x_{it}}}{1+e^{(\theta_i+\beta_t)}}$, and under the additional restriction $\beta_t = \beta$ for $t=1,\ldots,T$, we can solve for $\eta$ in terms of $\theta$ analytically, giving
\begin{align*}
\eta = \log \left[ (1+e^{\theta + \beta})^T -1 \right] = \eta(\theta,\beta) 
\end{align*} 
This function is monotone increasing in $\theta$, which is consistent with the basic intuition that higher ``catchability'' should be equivalent to smaller no-capture probabilities. In this case we can also give the condition
\begin{align*}
\theta_i > \log \left[ \left( \frac{1}{1-\alpha} \right)^{1/T}-1 \right]-\beta
\end{align*}
for individual $i$ to be $\alpha$-observable. If we further assume that $G^*$ has density $g^*$, we can give the induced distribution exactly
\begin{align}
f^*(\eta) = \left| \frac{1/T (1+e^{\eta})^{(1-T)/T}e^{\eta}}{(1+e^{\eta})^{1/T}-1} \right| g^*\left(\log \left[ (1+e^{\eta})^{1/T}-1 \right] -\beta \right). \label{eq:detamth}
\end{align}
When $|\eta|$ is large, the Jacobian factor in (\ref{eq:detamth}) behaves like a constant, and the term $\log[(1+e^{\eta})^{1/T}-1] \approx \eta$. So the tails of $f^*$ behave much like the tails of $g^*$.

\subsection{Additional Simulation Studies}
\begin{figure}[h]
 \centering
 \includegraphics[width=0.5\textwidth]{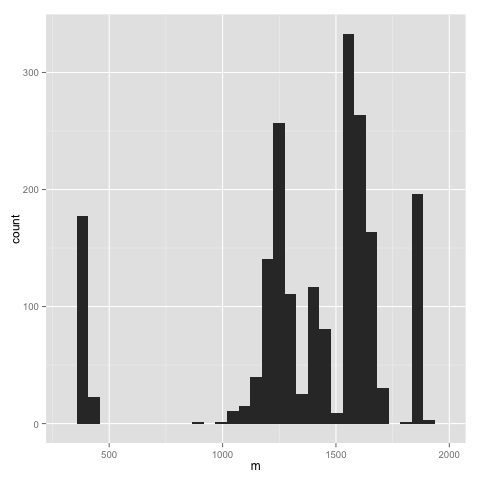}
 \caption{Distribution of values of $m$ in the simulation studies} \label{fig:mdist}
\end{figure}

In this section, we provide results for six additional simulation studies -- scenarios 4-9 in Table \ref{tab:simsumm} in the Appendix. The simulations are constructed as described in \S \ref{sec:simsetup}, and the choice of $G^*$ for data generation is varied across each simulation. The six additional cases considered here are mostly mixture models, though we also consider an atomic case and one in which the data generating model is the indirect Gamma. Figure \ref{fig:mdist} shows the distribution of observed population size $m$ across the 200 replicates of all nine simulations. In most cases, $m \in [1000,2000]$. Recall that in every simulation, the total population size $N=2000$.

\begin{figure}[h]
\begin{tabular}{ccc}
\includegraphics[width=0.3\textwidth]{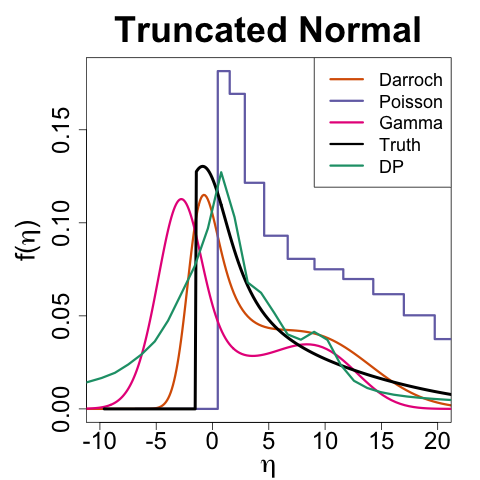} & \includegraphics[width=0.3\textwidth]{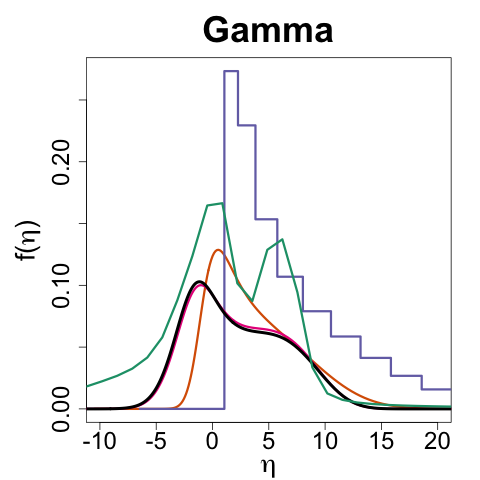} & \includegraphics[width=0.3\textwidth]{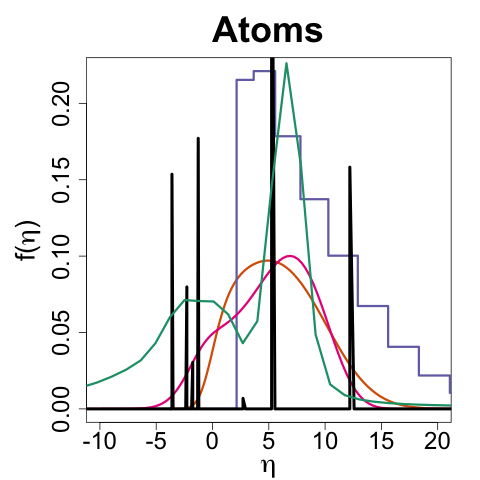} \\
\includegraphics[width=0.3\textwidth]{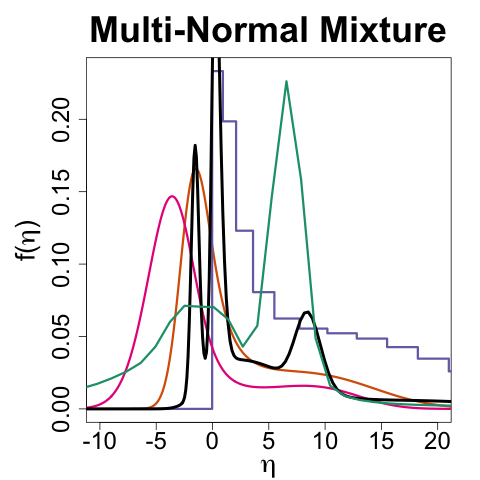} & \includegraphics[width=0.3\textwidth]{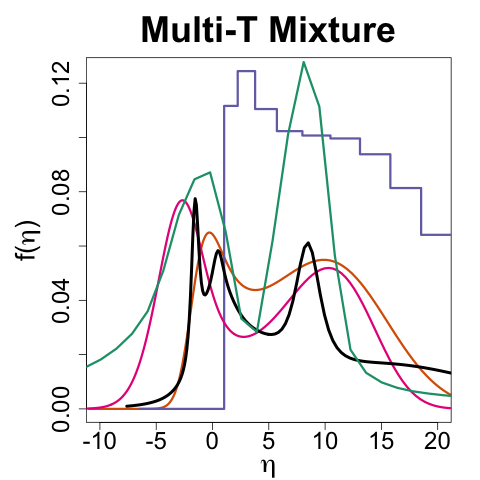} & \includegraphics[width=0.3\textwidth]{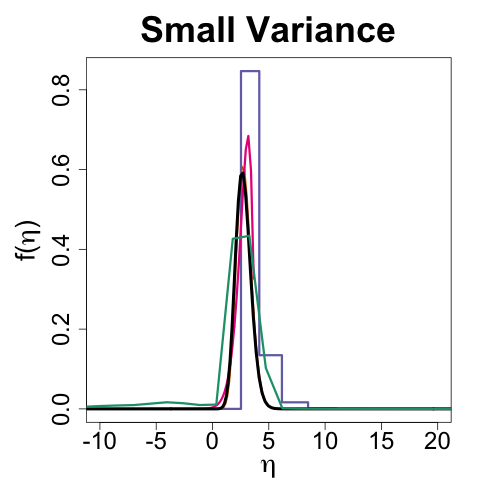} \\
\end{tabular}
\caption{Estimated density $f^*$ of observability $\eta$ on the logit scale in the four models considered, plotted with the true $f^*(\eta)$, for six of the ten simulation studies } \label{fig:etas2}
\end{figure}

Shown in Fig. \ref{fig:etas2} are estimates $\widehat{f^*}$ of the density $f^*(\eta)$ on the logit scale induced by the estimated distribution of the random effects $\widehat{G^*}$. The Dirichlet process model is generally successful at identifying the locations of significant modes, while the parametric models often miss one or more of the modes.   

\begin{figure}[h]
\centering
\begin{tabular}{ccc}
\includegraphics[width=0.3\textwidth]{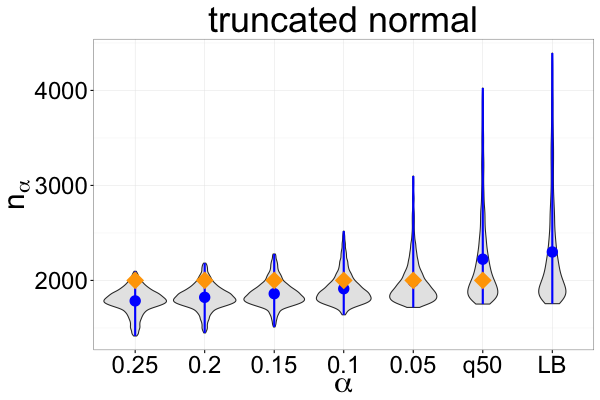} & \includegraphics[width=0.3\textwidth]{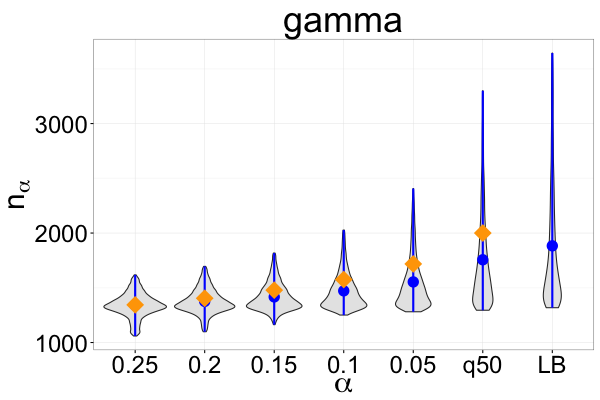}  & \includegraphics[width=0.3\textwidth]{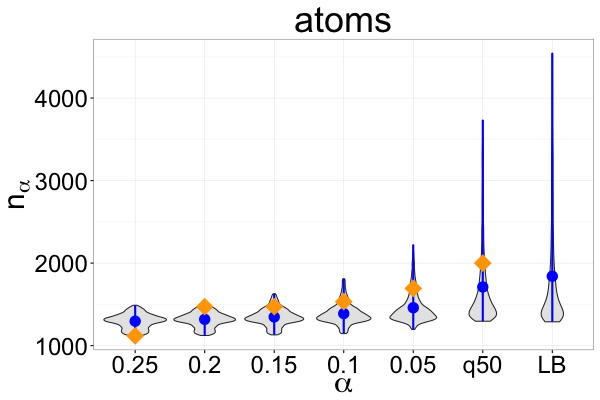} \\
\includegraphics[width=0.3\textwidth]{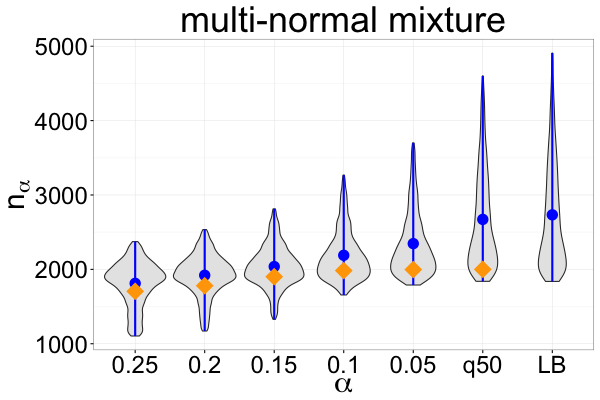} & \includegraphics[width=0.3\textwidth]{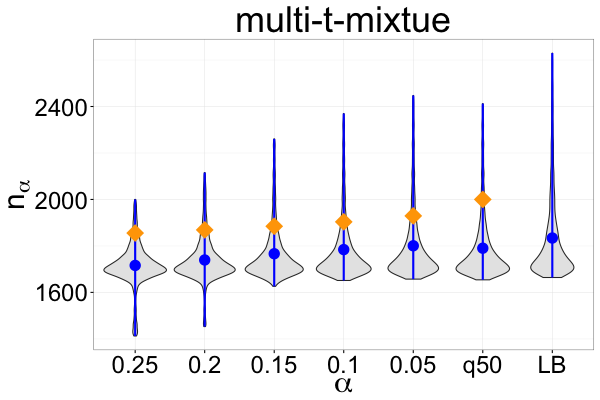} & \includegraphics[width=0.3\textwidth]{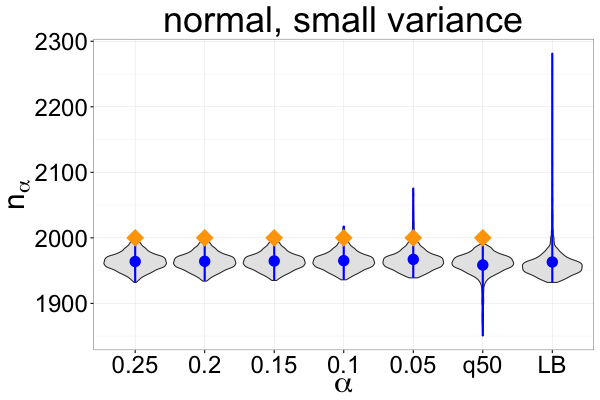} \\
\end{tabular}
\caption{Estimates of the posterior distribution for $N_{\alpha}$ for several $\alpha$ values for the Dirichlet process model; also shown are results for $\alpha  = \alpha_{\inf,0.5}$. The blue dots correspond to posterior means, the yellow dots to the true value of $N_{\alpha}$, and the blue lines to a 95 percent credible interval. } \label{fig:nalphas2}
\end{figure}

Figure \ref{fig:nalphas2} shows summaries of the estimated distribution of $N_{\alpha}$ for a single simulation replicate for a range of values of $\alpha$, as well as $\alpha_{\inf,0.5}$. The results correspond to simulation scenarios 4 through 9 from Table \ref{tab:simsumm} in the Appendix. The results are similar to those shown in the main text for simulation scenarios 1 through 3. As the value of $\alpha$ decreases, posterior uncertainty about $\alpha$ increases. However, any nonzero value of $\alpha$ allows for much smaller credible regions than those for estimates of $N$. This is expected and appropriate, since the value of $N$ cannot be accurately estimated in finite samples. Additionally, point estimates of $N_{\alpha}$ are relatively accurate and fall within the credible regions. 

\ifx\dmvdraft\undefined
\begin{table}[ht]
\centering
\begin{tabular}{rrrrrrr}
  \hline
 & 0.25 & 0.2 & 0.15 & 0.1 & 0.05 & 0 \\ 
  \hline
darroch & 1.00 & 1.00 & 1.00 & 1.00 & 1.00 & 1.00 \\ 
  normal & 1.00 & 1.00 & 1.00 & 1.00 & 1.00 & 1.00 \\ 
  2 normal mixture & 1.00 & 0.98 & 1.00 & 1.00 & 1.00 & 0.99 \\ 
  truncated normal & 1.00 & 1.00 & 1.00 & 1.00 & 1.00 & 0.99 \\ 
  gamma & 1.00 & 1.00 & 1.00 & 1.00 & 1.00 & 1.00 \\ 
  atoms & 0.54 & 0.91 & 0.98 & 0.99 & 0.99 & 1.00 \\ 
  multi-normal mixture & 1.00 & 1.00 & 1.00 & 1.00 & 1.00 & 0.97 \\ 
  multi-t-mixtue & 1.00 & 1.00 & 1.00 & 1.00 & 1.00 & 0.99 \\ 
  normal, small variance & 0.92 & 0.93 & 0.94 & 0.95 & 0.97 & 0.96 \\ 
  eco example & 1.00 & 0.99 & 0.99 & 0.99 & 1.00 & 1.00 \\ 
   \hline
\end{tabular}
\caption{Coverage of 95\% intervals for $N_{\alpha}$ for several values of $\alpha$. } 
\label{tab:nalpha_cov}
\end{table}

\fi
\end{appendix}

\end{document}